\documentclass[%
 aip,
 amsmath,amssymb,
reprint,%
]{revtex4-1}
\usepackage{float} 
\usepackage{graphicx}
\usepackage{dcolumn}
\usepackage{bm}

\usepackage[utf8]{inputenc}
\usepackage[T1]{fontenc}
\usepackage{mathptmx}
\usepackage{etoolbox}

\makeatletter
\def\@email#1#2{%
 \endgroup
 \patchcmd{\titleblock@produce}
  {\frontmatter@RRAPformat}
  {\frontmatter@RRAPformat{\produce@RRAP{*#1\href{mailto:#2}{#2}}}\frontmatter@RRAPformat}
  {}{}
}%
\makeatother
\begin{document}

\preprint{AIP/123-QED}


\title{Ultrashort pulsed laser atmospheric filament properties and microwave
radiation inferred from S-band guided wave interaction and self-emission}
\author{E. L. Ruden}
\author{J. E. Wymer}
\author{J. A. Elle}
\author{A. C. Englesbe}
\author{A. P. Lucero}
\author{E. A. Thornton}
\author{A. Schmitt-Sody}
\affiliation{Air Force Research Laboratory, Directed Energy Directorate}

\begin{abstract}
The electrical conductivity $\sigma$ of the plasma filament left behind by an
ultrashort pulsed laser (USPL) optical pulse after it is geometrically then
self-focused in air via the Kerr effect is measured by attenuation of a
$3.2$~GHz TE$_{10}$ mode within an S-band waveguide through which the filament
passes, taking into account the characteristic radius $R$ of the filament, as
determined by fast camera visible light imaging. Models of the major air
constituents' ionization rate $W_{i}$ vs.\ local laser intensity $I$, and of
electron temperature $T$ and mean axial electron momentum $\left\langle
p_{{z}}\right\rangle $ vs.\ peak laser intensity $I_{0}$ are then used to
infer the filament's $I_{0}$, $T$, major species particle densities, current
decay rate $\nu$, and current time integral $Q$ after pulse passage. $Q$ is
independently measured via the filament's self-emission in the waveguide
signal for comparison. The theoretical far field microwave radiation pattern
for the later $Q=Q\left(  z\right)  $ is calculated. It compares favorably to
published measurements, provided $\nu$ is reduced to the value where the total
radiated power is equal to the center-of-mass electron kinetic energy imparted
by the laser's magnetic Lorentz force. Results are presented for a $30$~cm
long filament at a broad range of atmospheric pressures.

\end{abstract}
\volumeyear{year}
\volumenumber{number}
\issuenumber{number}
\eid{identifier}
\date[Date text]{date}
\received[Received text]{date}

\revised[Revised text]{date}

\accepted[Accepted text]{date}

\published[Published text]{date}

\startpage{1}
\endpage{2}
\maketitle

\section{Introduction}

Microwave radiation from a plasma filament formed in air by geometrical
convergence and subsequent Kerr focusing \cite{Sprangle14} of an NIR
ultrashort pulsed laser (USPL) has been observed \cite{Englesbe18}%
\cite{Englesbe21}\cite{Janicek20}\cite{Thornton24} and simulated
\cite{Sprangle04}\cite{Garrett21}\cite{Garrett25}. The radiation pattern is
found to be rotationally symmetric about the laser propagation ($z$) axis,
implying an axial current source. The goal of this paper is to infer local and
far field properties based on measurements of the characteristic radius $R$
(by visible light imaging), electrical conductivity $\sigma$ (by interaction
of the filament with the TE$_{10}$ mode of an S-band
waveguide\ \cite{Papeer11}), and the total time integral $Q$ of the filament's
axial current (from filament self-emission within the waveguide) over a wide
range of background pressures $p$ and distances along such a filament $z$. A
description of the diagnostic and its calibration are provided in Appx.~A.

The filamentation process is unstable \cite{Mechain04}, resulting in the laser
intensity within to be difficult to determine. The relationship $\sigma
=\sigma\left(  n_{1},n_{2},T\right)  $ (assembled from several references
\cite{Bittencourt04}\cite{Kawaguchi25}\cite{Kawaguchi21}\cite{Spitzer53}%
\cite{Viegas71a}) depends on electron temperature $T$ and the densities
$n_{i}$ ($i=1,2$) of electrons released from O$_{2}$ and N$_{2}$,
respectively, for a given initial neutral density $n_{0}$ (single ionization
and net charge neutrality are assumed). In conjunction with the ionization
rates $W_{i}=W_{i}\left(  I\right)  $ of O$_{2}$ and N$_{2}$
\cite{Ruden25Qmodn} vs.\ local instantaneous intensity $I$ and electron
temperature $T=T\left(  I_{0}\right)  $ vs. peak intensity $I_{0}$ upon
electron thermalization \cite{Ruden25QmodT}, we find $I_{0}$, $n_{i}$, and $T$
along the filament from the $\sigma$ measurements by inversion. $I=I\left(
t\right)  $ is assumed to have a Gaussian time dependence with the measured
pulse width for this.

The $T\left(  I_{0}\right)  $ model per se is well-validated for our present
application since it amounts to a local fit to published electron kinetic
energy spectra for $I_{0}\sim10^{14}$~W/cm$^{2}$ of interest here. It assumes,
though, that the high energies actually measured deviate from the first order
Strong Field Approximation (generalized in the reference's \cite{Ruden25QmodT}%
Appx.~A) due to electrons rescattering off their parent ion (a second order
effect). The resulting magnetic Lorentz force before and after rescattering,
then, provides an estimate of mean \emph{post-optical} axial electron momentum
$\left\langle p_{{z}}\right\rangle $ vs.\ $I_{0}$. \textquotedblleft
Post-optical\textquotedblright\ here refers to time when the pulse has past,
but the axial current has not had time to Ohmically decay to a significant
degree. An analytic estimate $\nu_{1}$ of the subsequent current decay rate
$\nu$ of a \emph{hypothetical} steady state filament is found under the
assumption that the current remains well-distributed throughout the cross
section during decay. By \textquotedblleft steady state\textquotedblright, we
mean solutions with $z$ and $t$ having only a $z^{\prime}=z-ct$ dependence,
where $c$ is the speed of light. The ratio $Q=I_{\text{f}}/\nu_{1}$, then, is
an estimate of $Q$ for a steady-state filament, where $I_{\text{f}}$ is the
current after the electron's post-optical kinetic energy per unit length
$U_{0}$ (based on $\left\langle p_{{z}}\right\rangle $) has been minimally
transferred to the EM field that results from it. An independent direct
measurement of $Q$ based on filament self-emission is presented for
comparison. The latter $Q$, in addition to the $\sigma$ and $R$ measurements,
provide a estimate of steady state $\nu=\nu_{0}$ under the complementary
assumption that $I_{\text{f}}$ is driven to the surface during the energy
conversion process, and the associated EM field diffuses back in during decay.
The two estimates are in reasonable agreement despite the fact that $\nu$,
based on the following, is generally much lower.%

\begin{figure}[H]\includegraphics[width=3.375in]{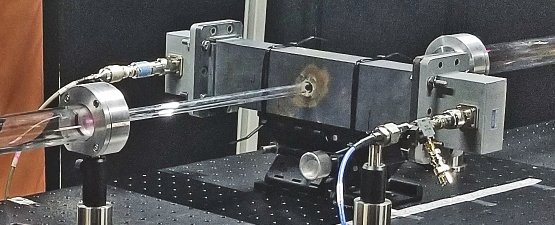}\end{figure}           

\textbf{Fig.~1 }Photograph of experimental hardware. Laser beam enters from
the left within joined fused quartz tubes, with the small one passing through
the diagnostic waveguide.

\bigskip

The far field radiation is modeled as a collective effect of the entire
filament with principal frequency $f_{1}=c/\left(  2L(1-\cos\theta)\right)  $
for which (unlike in steady state) $Q=Q\left(  z\right)  $. $L$ , here, is the
filament length and $\theta$ is emission angle. The total energy radiated due
to the directly measured $Q\left(  z\right)  $ far exceeds the laser-imparted
center of mass electron kinetic energy $K_{0}$ available to produce it for
either $\nu=\nu_{0}$ or $\nu=\nu_{1}$, except for the lowest $p$. A much lower
upper bound on $\nu=\nu_{\max}$, which assumes 100\% conversion, results in a
radiation pattern similar to published measurements \cite{Englesbe18} taken
with the same equipment and geometry, albeit with a laser pulse energy of
$40$~mJ instead of our $30$~mJ.

A Ti:sapphire USPL creates linearly polarized $30$~mJ optical pulses at a
repetition rate of $f_{\text{r}}=10$~s$^{-1}$ (for signal averaging purposes)
with a center wavelength of $800$~nm.\ A single-shot second order
autocorrelator measures the laser's full width at half maximum (FWHM) pulse
length to be $50$~fs. A $50$~mm diameter beam is geometrically focused in air
at various values of $p$ by a $300$~cm focal length spherical mirror. The
measurements are taken along a range of distances $z$ from the focusing mirror
after the beam has subsequently undergone further self-focusing by the Kerr
effect \cite{Sprangle14}, resulting in multiple plasma filaments. Data
analysis assumes a single filament, so is only accurate at $z$ locations where
with one dominant filament. To control $p$, the filament is formed within a
$70$ cm long fused quartz tube of inner radius $R_{1}=3.92$~mm and outer
radius $R_{2}=6.15$~mm. The tube passes orthogonally across the narrow width
(TE direction) through tight-fitting holes through the center of a $41.3$~cm
long WR284 ($3.4$~cm by $7.2$~cm) Cu S-band waveguide with $R_{0}=50$~$\Omega$
coax adapters on opposite ends, as pictured in Fig.~1. The laser is polarized
orthogonal to the TE direction.

The interaction of the filament with a $3.2$~GHz TE$_{10}$ mode driven at one
end of the waveguide and received at the other has both resistive
(dissipative) and inductive (reactive) contributions. To interpret the signal,
a calibration table of unit length conductance $G=\pi R^{2}\sigma$ of a round
cylindrical filament of uniform conductivity $\sigma$ vs.\ radius $R$ and
TE$_{10}$ power attenuation factor $A$ is created based on numerous 3-D
continuous wave simulations determining $A$ for a wide range of $\sigma$ an
$R$ values. Fast framing camera images are used to measure the filament's
radial luminosity profile. The images, though, imply the actual filament
conductivity profile $\sigma\left(  r\right)  $, where $r$ is the cylindrical
radius coordinate, is better approximated by a Gaussian. Conversion factors
between the uniform cylinder parameters $G/\left(  2\pi R\right)  $ and $R$ of
table and a Gaussian profile's $\sigma\left(  0\right)  $ and $s$ are derived
based on them sharing the same unit length conductance and inductance.

The filament produces its own signal superimposed on the attenuated TE$_{10}$
signal that is measured separately in isolation and subtracted to determine
$A$. This self-emission signal is also used to measure $Q$. To calibrate it, a
unipolar current pulse with a Gaussian time history with a short standard
deviation ($5$~ps) relative to S-band is sent through the filament in the
direction of the laser pulse in a time domain simulation of the waveguide
geometry similar to that described above, but with no TE$_{10}$ driver. It
results in a simulated receiver voltage $V_{\text{R}}$ time history similar to
the experimental one. The ratio of the simulated and experimental signals is
used to determine the charge $Q$ associated with a given signal $V_{\text{D}}$
of a power detector attached to the receiver output used to measure $A$.

\section{TE$_{10}$ attenuation factor $A$ measurements}

The waveguide diagnostic design and simulation are the subjects of Appx.~A.
The coax adapter at one end of the waveguide is used as a receiver (R) to
record output power $P_{\text{R}}$ vs.\ time $t$ averaged over $50$ shots with
and without a filament and with and without $3.2$~GHz TE$_{10}$ mode
excitation coming from a transmitter with input impedance $R_{0}$ attached to
an identical adapter at the other end. There is a significant signal when a
filament is present, but with the TE$_{10}$ drive off, due to filament
self-emission. This is subtracted from the signal with the drive on to isolate
the signal that results from TE$_{10}$ mode attenuation\ due to the filament.

The input to a Hewlett Packard 8470B crystal detector \cite{HP8470B} with
input impedance $R_{0}$ and nominal output shunt capacitance $C_{0}=30$~pF is
attached to the receiver output for the power measurement. It feeds a $50$
$\Omega$ coax cable connected to a digital oscilloscope with input impedance
$R_{0}$, $13$~GHz bandwidth, and $80\times10^{9}$~s$^{-1}$ sample rate. The
detector (D) has an output voltage $V_{\text{D}}$ proportional to the negative
of the receiver's CW power $P_{\text{R}}$, with nominal sensitivity $K=0.5$
mV/$\mu$W in the $10$~MHz - $8$~GHz range.%
\begin{equation}
V_{\text{D}}=-KP_{\text{R}}=-K\frac{\left\langle V_{\text{R}}^{2}\right\rangle
\text{ }}{R_{0}}\text{ } \label{X1}%
\end{equation}
where $V_{\text{R}}$ is the dynamic (oscillating) receiver output voltage, and
\textquotedblleft$\left\langle {}\right\rangle $\textquotedblright\ represents
the time-dependent cycle-average of the enclosed. The cycle-averaging time
window corresponds to the nominal detector response time of $R_{0}C_{0}=1.5$
ns (several cycles for S-band).

Figure~2 plots the power vs.\ time recorded by the HP detector due to
self-emission, based on Eq.~\ref{X1}, for a range of background air pressures
$p$. Multiple reflections of the high frequency / high order modes above
S-band (Fig.~12), for which the receiver is poorly matched, contribute
significantly to its long decay time.%

\begin{figure}[H]\includegraphics[width=3.375in]{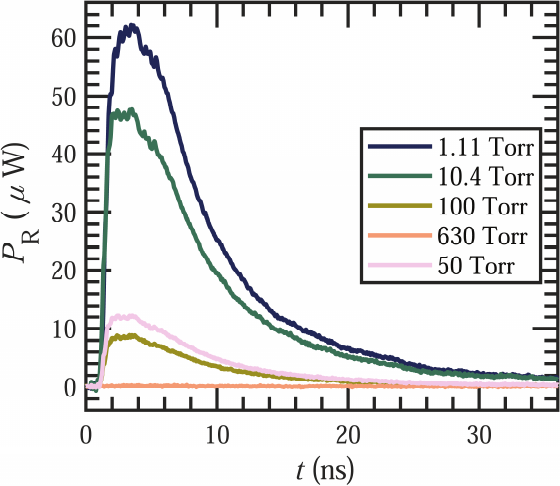}\end{figure}           

\textbf{Fig.~2 }HP 8470B crystal detector signal averages\ due to filament
self-emission, calibrated for power from the S-band waveguide receiver for a
range of background pressures $p$ at filament range from the focusing mirror
$z=307$~cm (near the filament center). Though not a pressure studied in detail
below, the self-emission signal for $p=50$~Torr at this $z$ is added for use
in conjunction with the data plotted in Fig.~12 to determine the time integral
$Q$ of the filament's self-current.%

\begin{figure}[H]\includegraphics[width=3.375in]{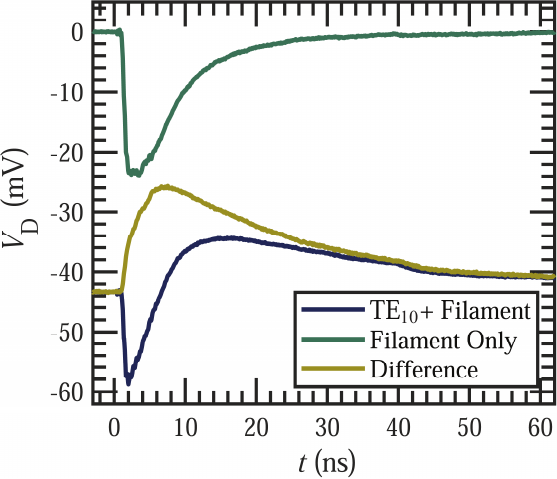}\end{figure}           

\textbf{Fig.~3} Uncalibrated HP crystal detector signal averages with a
filament and the $3.2$~GHz drive on (bottom), self-emission\ of the filament
only (top), and the difference (middle). Parameter $A$ is the peak of the
latter divided by its initial value. $p=10.4$~Torr with the waveguide at
$z=307$~cm is used for this example (near the center of the filament).%

\begin{figure}[H]\includegraphics[width=3.375in]{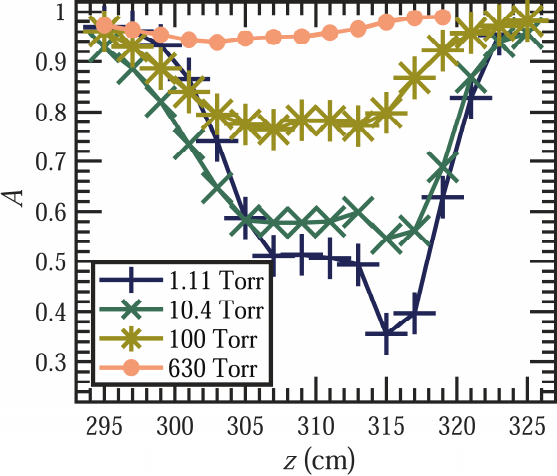}\end{figure}           

\textbf{Fig.~4} Attenuation parameter $A$ (corrected for self-emission) vs.
distance $z$ from the focusing mirror for a range of $p$ values.

\bigskip

Parameter $A$ is the factor by which the HP\ power detector signal
$V_{\text{D}}$ is reduced in peak amplitude due to absorption and reflection
of the TE$_{10}$ mode transmitted from the other end of the waveguide due to
the filament's resistive and inductive impedance. Correcting for self-emission
takes into account the random phase of the TE$_{10}$ mode when the laser
fires. The voltage waveform at the direct receiver output $V_{\text{R}}$ (sans
detector) is the sum of the voltage of the attenuated TE$_{10}$ mode $V_{1}$
and the self-emission voltage $V_{2}$. The former can be expressed as,%
\begin{equation}
V_{1}=V_{10}\sin\left(  \omega_{10}t+\phi+\phi_{0}\right)  \label{X2}%
\end{equation}
where $\omega_{10}$ is the mode's angular frequency, $\phi_{0}$ is the mode's
phase at the time of laser firing, and $V_{10}$ and $\phi$ are the receiver
voltage amplitude and phase shift of the mode, respectively, as modified by
the filament.

We have, then, for use in Eq.~\ref{X1},%
\begin{equation}%
\begin{tabular}
[c]{l}%
$\frac{V_{\text{R}}^{2}}{R_{0}}=\frac{\left(  V_{1}+V_{2}\right)  ^{2}}{R_{0}%
}=\frac{V_{1}^{2}}{R_{0}}+\frac{V_{2}^{2}}{R_{0}}$\\
$+2V_{2}V_{10}\sin\left(  \omega_{10}t+\phi+\phi_{0}\right)  $%
\end{tabular}
\ \ \ \label{X3}%
\end{equation}
The second line is the random interference term. $P_{\text{R}}$ with the
$3.2$~GHz drive on is observed to vary by $\sim10$\% from shot to shot due, in part
at least, to random variation in $\phi_{0}$. Our averaging the results of $50$
shots for each data point reduces it by an order of magnitude since the
interference term averages to zero. What remains is the sum of the power due
to the two contributions (attenuation and self-emission) in isolation. The
self-emission signal recorded separately (also averaged over 50 shots for
noise reduction) is subtracted to isolate the filament's effect on TE$_{10}$.
Figure~3 plots an example of such signals and their difference. $A$ is equated
to the ratio of the peak of average detector signal so corrected divided by
its value prior to filament formation.

$A$ vs.\ distance $z$ from the focusing mirror for a range of $p$ is plotted
in Fig.~4. $A$ is measured for 30 mJ shots for a range of $p$ and $z$ values.
$A$ is one of the two parameters needed to determine the filament's $\sigma$
from the filament's unit length conductance $G=G\left(  A,R\right)  $
calibration table (Fig.~21). This table is based on a filament whose
properties do not change with time, while the actual filament is formed
quickly and $G$ decays slowly relative to the detector response time. Figure~3
shows that the decay time is an order of magnitude longer than the detector
response time, so an underestimate of $G$ by $\sim10$\% is expected. No effort
is made to compensate for this since a larger error due to the minor side
filaments (which leads to an \emph{overestimate} of the presumed single
filament's $G$) is expected. The other parameter needed to determine $G$ is
the radius $R$ of a uniform cylinder with the same inductance per unit length
as that inferred from imaging.

\section{Radius $R$ found from images}%

\begin{figure}[H]\includegraphics[width=3.375in]{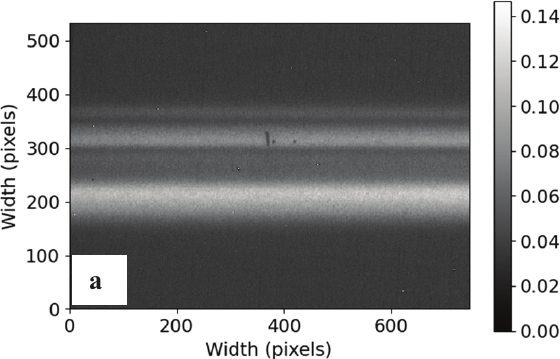}\end{figure}           
\begin{figure}[H]\includegraphics[width=3.375in]{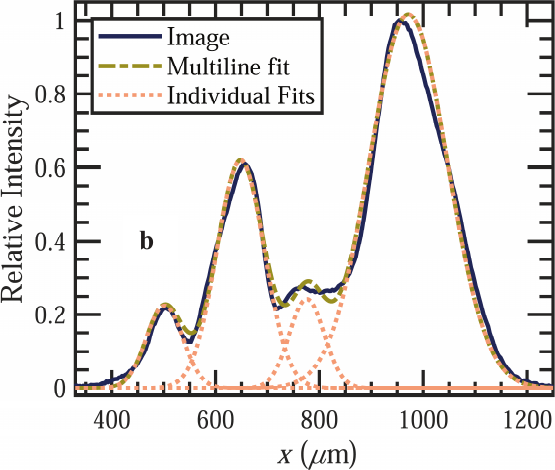}\end{figure}           

\textbf{Fig.~5 }Filament image at $z=307$~cm for $10.4$~Torr (a) and
corresponding plot\ of intensity vs.\ transverse distance $x$ showing several
filaments and their Gaussian fits (b).

\bigskip

Visible light imaging of the filament taken at a range of distances $z$ from
the focusing mirror provides the basis for estimating an effective value of
$R$ for use with the $G\left(  A,R\right)  $ calibration table to determine
$G$. A 4 Quik E-Dig from Stanford Computer Optics, Inc.\ with an S-20 spectral
response and $8.3$~$\mu$m square image (sensor) pixel size is used for this.
It is set to a gain $1000$ and a gate window of $2$ ns. The lens assembly is a
bellows-mounted Canon F2.8 macro photo lens with a focal length of $L_{0}=35$
mm stopped down to $F=5.6$. This results in the sharpest images with a depth
of field sufficient to well-resolve all filaments simultaneously. The
transverse object pixel size is $3.00\pm0.04$ $\mu$m based on a numerical
analysis of several 1951 USAF Resolution Test Chart images, implying
magnification $M=2.72.$ The chart is placed inside the quartz tube for this
measurement to take into account minor cylindrical lensing, and to assess the
tube's effect on resolution. General purpose resolution of isolated features
is roughly $s_{1}\approx15$ $\mu$m (5 pixels), based the distance required for
the calibration lines to make an intensity transition midway between lines
that is 50\% of the peak-to-peak range.

Figure~5 is an example of a filament image and its intensity vs.\ transverse
distance $x$ averaged over the range of $z$ for the image at $z=307$. It shows
multiple filaments formed at $p=10.4$~Torr, where the multifilamentation is
significant. Such images are highly reproducible, implying the filamentation
pattern is determined by aberration, such as from the angled spherical mirror.
The images used in this paper are the result of averaging over $50$ images to
reduce noise. A single image does not vary by an observable amount in the $z$
direction due to the small field of view, as shown. The filament is not
well-represented by the round cylinder of uniform conductivity of the
simulations. A Gaussian makes a better fit. Multiline Gaussian fits using
parameters iterated by hand are overlaid in Fig.~5b.

\newpage

\begin{figure}[H]\includegraphics[width=3.375in]{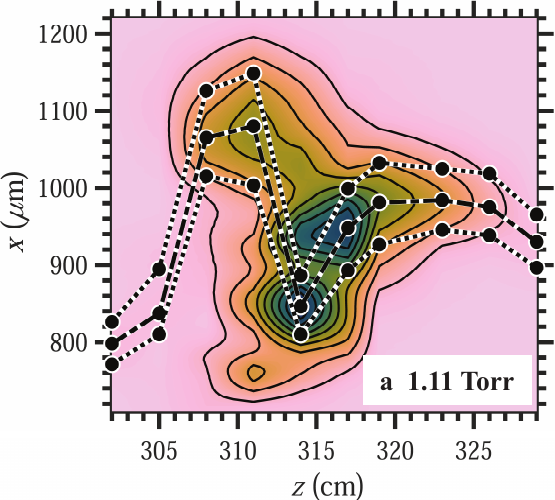}\end{figure}           
\begin{figure}[H]\includegraphics[width=3.375in]{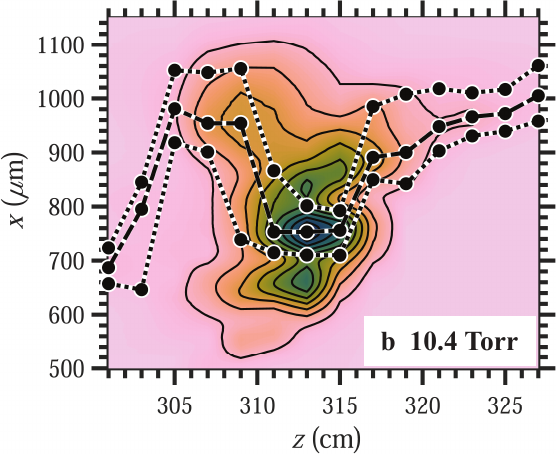}\end{figure}           
\begin{figure}[H]\includegraphics[width=3.375in]{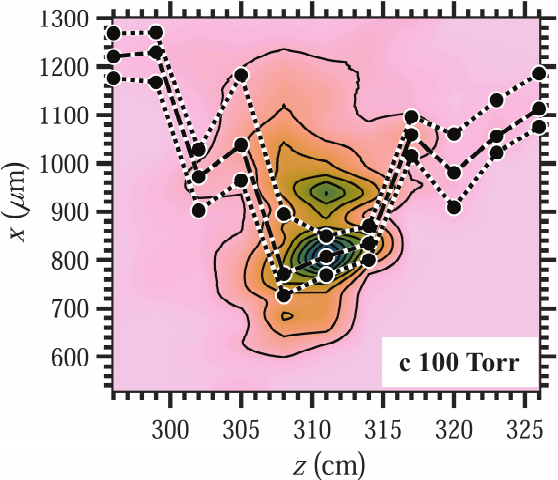}\end{figure}           
\begin{figure}[H]\includegraphics[width=3.375in]{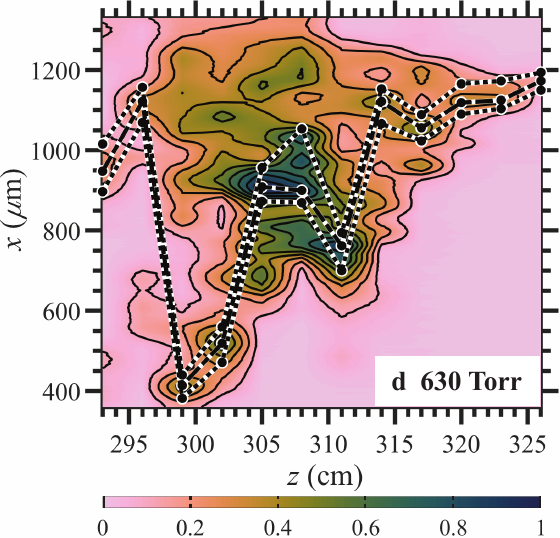}\end{figure}           

\textbf{Fig.~6 }Contour plots of composites of the image line-outs from at all
$z$ locations and all pressures $p$ studied. The solid line tracks the peak
intensity vs.\ $z$, with the dashed line above and below tracking the point at
which intensity falls by a factor of $\exp\left(  -1/2\right)  $ on either
side (one standard deviation $s$ of a Gaussian profile).%

\begin{figure}[H]\includegraphics[width=3.375in]{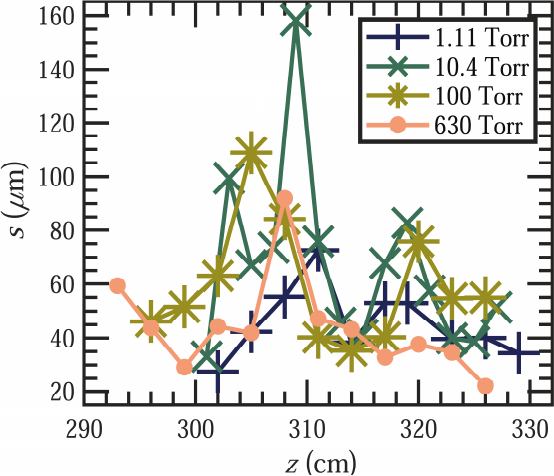}\end{figure}           

\textbf{Fig.~7 }Standard deviation $s$ estimated as the $\exp\left(
-1/2\right)  $ half-width of peak intensity at range $z$, based on the data
plotted in Fig.~6.

\bigskip

In general, we find the filament current to be well-diffused across the
filament's cross section in the simulations for the cases of interest. For
best use of the $G(A,R)$ calibration table, then, we take the simulations' $R$
at a given range from the mirror $z$ to be that of a cylinder with uniformly
distributed current density $J_{0}$ with the same inductance per unit length
as a filament with an axial Gaussian current density $J\left(  r\right)  $
distribution in cylindrical radius $r$ of standard deviation $s$,%
\begin{equation}
J\left(  r\right)  =J\left(  0\right)  \exp\left(  -\frac{r^{2}}{2s^{2}%
}\right)  \label{X5}%
\end{equation}
Figure~6 plots a composite of the line-outs from images at all $z$ locations
and all $p$. The peak intensity is plotted with the intensity at a presumed
Gaussian $+s$ and $-s$ distance above and below the peak also shown. $s$ for
each $z$ position is approximated as the image's inferred $\exp\left(
-1/2\right)  $ half-width. This assumes volume emissivity is proportional to
$J\left(  r\right)  $ for the purpose of calculating an inductively equivalent
$R$ (the imaged Abel projection of a Gaussian distribution is a Gaussian
distribution of the same width). Figure~7 plots $s$ vs.\ $z$.

To determine the relationship between the Gaussian distribution's $J\left(
0\right)  $ and the simulations' uniform cylindrical $J_{0}$ and $R$, we first
equate their total currents,%

\begin{equation}
2\pi s^{2}J\left(  0\right)  =\pi R^{2}J_{0} \label{X6}%
\end{equation}
To determine the relationship between $R$ and $s$ in Eq.~\ref{X6}, we equate
the two configurations' inductances by equating the magnetic energy per unit
length for a given current in the magnetostatic limit. Given Eq.~\ref{X5} and
Eq.~\ref{X6}, integrating magnetic energy density $B^{2}/\left(  2\mu
_{0}\right)  $ over area from $r=0$ to a finite value (for now) of $r=R_{+}$,
where $B$ for each configuration is based on the integrated form of Ampere's
law, and equating the results for the two configurations results in,%

\begin{equation}
\int_{0}^{R_{+}}\left(  1-\exp\left(  -\frac{r^{2}}{2s^{2}}\right)  \right)
^{2}\frac{dr}{r}=\frac{1}{4}+\ln\frac{R_{+}}{R} \label{X7}%
\end{equation}

The l.h.s.\ of Eq.~\ref{X7} is the normalized magnetic energy of the Gaussian
filament (with terms with units on both sides cancelling). The sum on the
r.h.s.\ are the contributions from inside and outside the uniform cylinder,
respectively. Upon expanding the term in parentheses squared and expressing
the integral as that from $r=0$ to $R$ plus that from $r=R$ to $R_{+}$, one
term in the later results in $\ln\left(  R_{+}/R\right)  $ upon integration.
This cancels the same expression on the right, leaving us with,%
\begin{equation}%
\begin{tabular}
[c]{l}%
$\int_{0}^{R/s}\left(  1+\exp\left(  -u^{2}\right)  -2\exp\left(  -\frac
{u^{2}}{2}\right)  \right)  \frac{du}{u}$\\
$+\int_{R/s}^{R_{+}/s}\left(  \exp\left(  -u^{2}\right)  -2\exp\left(
-\frac{1}{2}u^{2}\right)  \right)  \frac{du}{u}=\frac{1}{4}$%
\end{tabular}
\ \label{X8}%
\end{equation}
\newline where $u=r/s$. This eliminates the far field contribution common to
both geometries that diverges as $R_{+}\rightarrow\infty$. Taking now
$R_{+}\rightarrow\infty$,%

\begin{equation}
\frac{R}{s}=2\exp\left(  \frac{1}{4}-\frac{\gamma}{2}\right)
=1.924\,3\ \ \label{X9}%
\end{equation}
where $\gamma$ is the Euler--Mascheroni constant.

\section{Electrical conductivity $\sigma$ found from attenuation parameter $A$
and radius $R$, and other properties inferred}

Given $R$ from Eq.~\ref{X9}, $s$ vs.\ $z$ from Fig.~7, and $A$ vs.\ $z$ from
Fig.~4, $\sigma\left(  0\right)  $ vs.\ $z$ for the filament may be estimated
in terms of $G$ from the $G(A,R)$ calibration matrix plotted in of Fig.~21 by
assuming $\sigma\left(  r\right)  $ too has a Gaussian distribution with
standard deviation $s$,%
\begin{equation}
\sigma\left(  r\right)  =\sigma\left(  0\right)  \exp\left(  -\frac{r^{2}%
}{2s^{2}}\right)  \label{X10}%
\end{equation}
This is justified since it results in a uniform $\mathbf{E}$ field, based on
Ohm's law. Equating $G$ now to the Gaussian profile's own unit length
conductance, determined by integrating $\sigma\left(  r\right)  $ over the
cross sectional area, we have, as plotted in Fig.~8,
\begin{equation}
\sigma\left(  0\right)  =\frac{G}{2\pi s^{2}}\label{X11}%
\end{equation}

\begin{figure}[H]\includegraphics[width=3.375in]{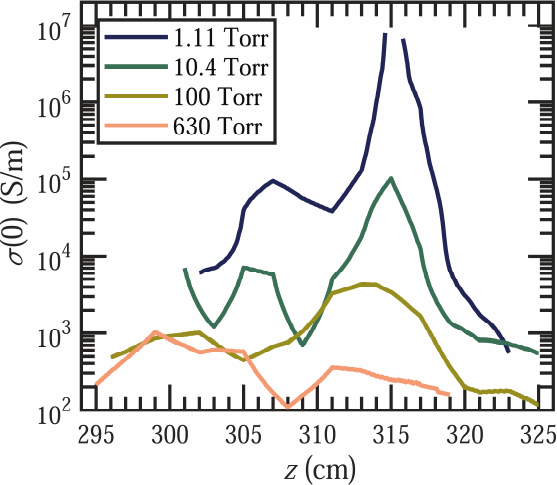}\end{figure}           

\textbf{Fig.~8 }Filament peak conductivity $\sigma\left(  0\right)  $ vs.\ $z$
of an assumed single filament vs.\ $z$ and $p$ (specified in legend) of an
assumed Gaussian profile with a width of standard deviation $s$ vs.\ $z$
plotted in Fig.~7 (Eq.~\ref{X11}). The surges in $\sigma\left(  0\right)  $
near $z=315$ cm for lower $p$ values are due to excessive violation of the
single-filament approximation.

\bigskip

The availability of monotonic expressions for the peak free electron densities
$n_{1}=n_{1}\left(  I_{0}\right)  $ released from O$_{2}$ and $n_{2}%
=n_{2}\left(  I_{0}\right)  $ released from N$_{2}$ and peak electron
temperature $T=T\left(  I_{0}\right)  $ all as functions of peak laser
intensity $I_{0}$ within the filament permit these properties (in addition to
$I_{0}$ itself) to be determined from the $\sigma$ measurements, given a model
of $\sigma=\sigma\left(  n_{1},n_{2},T\right)  $. The detailed steps are
specified in the caption to Fig.~10.

$n_{i}\left(  I_{0}\right)  $ for $i=1,2$ are found from the optical-cycle
averaged ionization rates' dependence on local intensity $I$ for air's main
constituents N$_{2}$ and O$_{2}$ \cite{Ruden25Qmodn} as the post-optical
solutions to $\partial n_{i}/\partial t=\left(  n_{i0}-n_{i}\right)
W_{i}\left(  I\right)  $, where $W_{i}\left(  I\right)  $ is the optical-cycle
averaged ionization rate ($\left\langle W\right\rangle _{\gamma}$ in this
reference), and $n_{i0}$ is the initial density of that species' neutral
molecule. For air, we take, $n_{10}=0.20\left(  p/p_{\text{a}}\right)
n_{\text{a}}$ and $n_{20}=0.80\left(  p/p_{\text{a}}\right)  n_{\text{a}}$,
where $p_{\text{a}}=760$~Torr and $n_{\text{a}}=2.687\times10^{25\text{ }}%
$m$^{-3}=$ Loschmidt's number \cite{NRL} (STP molecular air density). Assuming
a Gaussian time history for $I\left(  t\right)  $, the solution is,
\begin{equation}%
\begin{tabular}
[c]{l}%
$I=I_{0}\exp\left(  -\frac{t^{2}}{t_{\mathrm{L}}^{2}}\right)  $\\
$n_{i}=n_{i0}\left(  1-\exp\left(  -\int\nolimits_{-\infty}^{+\infty}%
W_{i}\left(  I\left(  t\right)  \right)  dt\right)  \right)  $%
\end{tabular}
\ \label{X12}%
\end{equation}
The USPL $e^{-1}$ half width $t_{\mathrm{L}}=30$~fs is determined from the
laser pulse full width at half maximum of $50$~fs divided by $2\sqrt
{\ln\left(  2\right)  }$. The results for each species are plotted in Fig.~9,
along with those of pulses of half and twice the pulse width of the data presented.%

\begin{figure}[H]\includegraphics[width=3.375in]{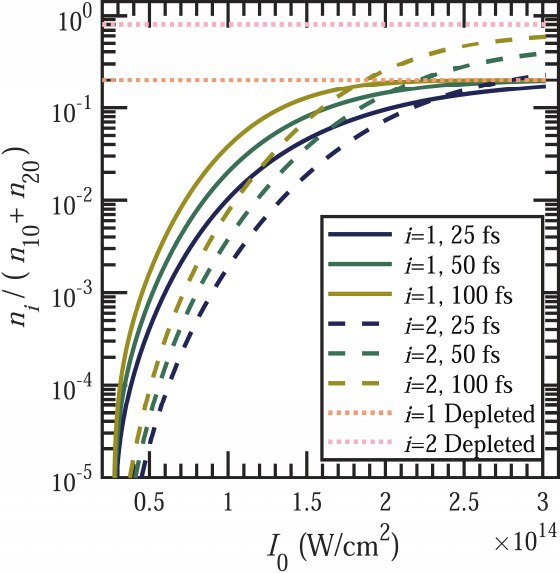}\end{figure}           

\textbf{Fig.~9 }$n_{1}/\left(  n_{10}+n_{20}\right)  $ (O$_{2}$) and
$n_{2}/\left(  n_{10}+n_{20}\right)  $ (N$_{2}$) vs.\ $I_{0}$ for a range of
USPL pulse widths in air with FWHM specified in the legend, based on Eqs.
\ref{X12}. Local ionization rates $W_{i}\left(  I\right)  $ for the two
species are determined by inversion of time-of-flight spectrometer
measurements of ion count \cite{Ruden25Qmodn}. The $50$~fs results are assumed
for this paper, with results for the other widths shown for reference.

\bigskip

$T\left(  I_{0}\right)  $ is taken to be that of electrons released from
O$_{2}$ alone \cite{Ruden25QmodT} ($T_{2}$ in ref.). This is justified since
$n_{2}$ (for N$_{2}$) only contributes 15\% to $n_{1}+n_{2}$ at $I_{0}%
=10^{14}$~W/cm$^{2}$, based on Eq.~\ref{X12}.2 (referring here to the equation
that starts on the second line of Eqs.~\ref{X12}), and has a temperature that
differs by only a few 10's of percent, based on similarity in N$_{2}$ electron
kinetic energy spectra exposed to $I_{0}$ values of this order
\cite{Okunishi08}\cite{Deng11}. In general, $T$ for USPL ionized electrons is
dominated by electron acceleration by the $\mathbf{E}$ field of the laser
subsequent to ionization, so a relatively weak dependence of $T$ on species is
expected. What difference there is, though, becomes more significant at higher
$I_{0}$, since $n_{2}$ rises much faster than $n_{1}$ beyond $I_{0}=10^{14}%
$~W/cm$^{2}$ (Fig.~9), becoming fully equal at $I_{0}=2.1\times10^{14}%
$~W/cm$^{2}$.

For our $\sigma\left(  n_{1},n_{2},T\right)  $ expression, consideration is
needed to take into account the sum \cite{Shkarofsky61}$^{:22\text{ }}%
$\cite{Gilardini72}$^{:2.6.31}$ of the electron momentum transfer collision
frequencies $\nu_{\text{c}}$ \cite{Shkarofsky61}$^{:2}$ vs.\ electron velocity
$v$ off of the neutral and ion species, and the 3.2~GHz S-band driver's
angular frequency $\omega_{10}=\allowbreak2.01\times10^{10}\allowbreak$
s$^{-1}$. Numbers following the colon in a reference number here and elsewhere
refer to the equation number within the reference itself. $\omega_{10}$
introduces an imaginary component to conductivity that our calibration
(Fig.~21) does not take into account. Fortunately, we only need the real
component $\sigma$ of conductivity to account TE$_{10}$ attenuation from Ohmic
heating \cite{Lieberman05}$^{:4.2.26}$ $\sigma E^{2}=\operatorname{Re}\left(
\mathbf{J}^{\ast}\cdot\mathbf{E}\right)  /2$, where $\mathbf{J}$ and and
$\mathbf{E}$ here are the complex current density and electric field vectors
at frequency $\omega_{10}$, and $E$ is the latter's magnitude. We may
reasonably assume the imaginary conductivity component has a negligible effect
on the filament's reactive impedance compared to the filament's inductance
(that \emph{is} accounted for), since the filament has a very small radius
relative to the waveguide width. Also, $\omega_{10}$ is found to be a factor
of $10$ smaller than the plasma frequency $\omega_{\text{p}}=\left(  \left(
n_{1}+n_{2}\right)  e^{2}/\left(  m\epsilon_{0}\right)  \right)  ^{1/2}$ for
$T>2.2$ eV for all cases considered, except for $p=1.1$~Torr, where $T>2.5$ is
needed, and our final results will all have a higher $T$, so consideration of
Drude conductivity \cite{Adamyan09} due to displacement current is also
omitted \emph{in this section}. To this end, the real component of $\sigma$
assumed is \cite{Bittencourt04}$^{:22.2.25}$,%
\begin{equation}%
\begin{tabular}
[c]{l}%
$\sigma=\frac{e^{2}\left(  n_{1}+n_{2}\right)  }{m\nu_{\text{c,eff}}}$
$\ \ \ \ \ \ \ \ \ \ \ \ \ \ \ \nu_{\text{c}}=\sum\limits_{i=1}^{3}%
\nu_{\text{c}i}$\\
$\frac{1}{\nu_{\text{c,eff}}}=\frac{8}{3\sqrt{\pi}}\left(  \frac
{m}{2k_{\text{B}}T}\right)  ^{5/2}\int_{0}^{\infty}\frac{v^{4}\nu_{\text{c}}%
}{\nu_{\text{c}}^{2}+\omega^{2}}\exp\left(  -\frac{mv^{2}}{2k_{\text{B}}%
T}\right)  dv$\\
$\nu_{\text{c}i}=v\left(  n_{i0}-n_{i}\right)  q_{i}\left(  v\right)  $,
$i=1,2$\\
$\nu_{\text{c}3}=v\left(  n_{1}+n_{2}\right)  q_{3}\left(  v\right)  $ \ \
\end{tabular}
\ \ \ \ \label{X13}%
\end{equation}
where $e$ is fundamental charge, $m$ is electron mass, $k_{\text{B}}$ the
Boltzmann constant, and $\omega=\omega_{10}$ for the S-band conductivity measurement.

$\nu_{\text{c}1}$, $\nu_{\text{c}2}$, and $\nu_{\text{c}3}$, in Eqs.~\ref{X13}
are the contributions to $\nu_{\text{c}}$ for electron scatter from O$_{2}$,
N$_{2}$, and singly charged positive ions of either species, respectively,
taking into account their dynamically changing densities due to ionization
(Eq.~\ref{X12}.2). $q_{i}\left(  v\right)  $ are the momentum transfer
collision cross sections using Kawaguchi \cite{LXCat_Kawaguchi} data for $i=1$
(O$_{2}$, plotted as $q_{\text{vm}}\left(  v\right)  $ in Fig.~2 of Kawaguchi
et al.\ 2025 \cite{Kawaguchi25}) and LISBON \cite{LXCat_LISBON} data for $i=2$
(N$_{2}$, plotted in Fig.~2 of Kawaguchi et al.\ 2021 \cite{Kawaguchi21}).

$q_{3}\left(  v\right)  $ is based on the DC conductivity $\sigma_{\text{SH}}$
of Spitzer and H\"{a}rm \cite{Spitzer53}$^{:33}$ for a\emph{ fully ionized
}plasma with adjustments to the Coulomb logarithm based on Lee and More
\cite{Lee84} and second order correction of Viegas \cite{Viegas71a}$^{:5}$,
\begin{equation}%
\begin{tabular}
[c]{l}%
$\sigma_{\text{SHLMV}}=\gamma_{E}\gamma_{\text{U1}}^{\text{e}}\left(
2\right)  \frac{32\sqrt{\pi}\epsilon_{0}^{2}m}{e^{2}\mathcal{L}}\left(
\frac{2k_{\text{B}}T}{m}\right)  ^{3/2}$ \ \ $\gamma_{E}=0.5816$\\
$\mathcal{L}=\ln\left(  \frac{b_{\max}}{b_{\min}}\right)  $ $\ b_{\max}%
=\max\left(  \sqrt{\frac{\epsilon_{0}k_{\text{B}}T}{\left(  n_{1}%
+n_{2}\right)  e^{2}}},\frac{1}{\left(  n_{1}+n_{2}\right)  ^{1/3}}\right)
$\\
$b_{\min}=\max\left(  \frac{e^{2}}{12\pi\epsilon_{0}k_{\text{B}}T},\frac
{\pi\hbar}{\sqrt{3mk_{\text{B}}T}}\right)  $\\
$\gamma_{\text{U1}}^{\text{e}}\left(  2\right)  =\frac{3\pi\mathcal{L}%
}{32\gamma_{E}}\left(  \mathcal{L}+E_{11}-\frac{D_{21}^{2}\left(
\mathcal{L}+E_{21}\right)  ^{2}}{B_{22}\left(  \mathcal{L}+C_{22}\right)
+D_{22}\left(  \mathcal{L}+E_{22}\right)  }\right)  ^{-1}$\\
$B_{22}=\sqrt{2}$\ $\ C_{22}=-0.4478$ \ $D_{21}=\frac{3}{2}$\ $\ D_{22}%
=\frac{13}{4}$\\
$E_{11}=-1.3670$ \ $E_{21}=-2.0337$ \ $E_{22}=-1.9824$%
\end{tabular}
\ \label{X14}%
\end{equation}
where $\epsilon_{0}$ is free space permeability.%

\begin{figure}[H]\includegraphics[width=3.375in]{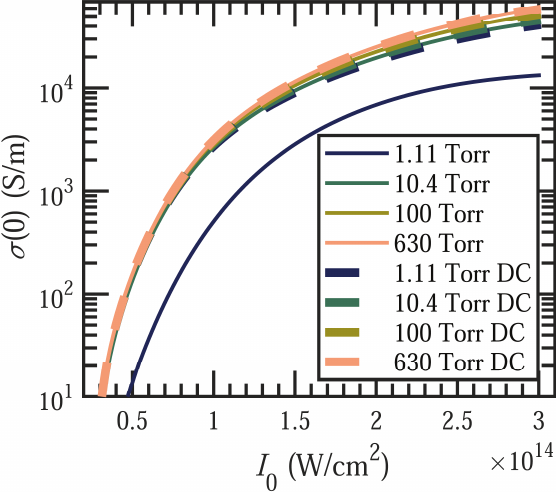}\end{figure}           

\textbf{Fig.~10} $\ $Plot of $\sigma\left(  0\right)  $ of Eq.~\ref{X11}
($\sigma$ on the filament axis), as approximated by $\sigma\left(
n_{1}\left(  I_{0}\right)  ,n_{2}\left(  I_{0}\right)  ,T\left(  I_{0}\right)
\right)  $, where $\sigma\left(  n_{1},n_{2},T\right)  $ is the solution to
Eq.~\ref{X13}.1, $n_{1}\left(  I_{0}\right)  $ and $n_{2}\left(  I_{0}\right)
$ are solutions to Eqs.~\ref{X12}, and $T\left(  I_{0}\right)  $ is $T_{2}$ in
reference \cite{Ruden25QmodT}. The solid lines are $\sigma\left(  0\right)  $
for an $\mathbf{E}$ field with the angular frequency of the S-band driver
$\omega=\omega_{10}$. $\sigma\left(  0\right)  $ for $\omega=0$ (dashed lines)
are shown for reference.

\bigskip

Lee and More's replacement of $T$ in the first choice (Debye length
$\lambda_{\text{D}}$) for (impact parameter cutoff) $b_{\max}$ with $\left(
T^{2}+T_{\text{F}}^{2}\right)  ^{1/2}$ and/or more detailed degeneracy
considerations are omitted since the lowest $T$ for our problem is $2.05$ eV
while the highest (Fermi temperature \cite{Kittel80}$^{:7.8}$) $T_{\text{F}}$
is $2.0\times10^{-2}\allowbreak$ eV, occurring at $p=630$~Torr where air is
58\% ionized at the highest $I_{0}$, as shown for the $50$~fs plots of Fig.~9.
The second choice (mean interionic separation) for $b_{\max}$ is only needed
near this highest $p$ and lowest $T$ considered, where its value is $4.3$ nm
(vs.\ $2.8$ nm for $\lambda_{\text{D}}$). The first choice (impact parameter
for 90$^{\circ}$ elastic scatter of an electron with the mean thermal velocity
\cite{Goldston95}$^{:11.17}$) for $b_{\min}$ is always greater than the second
(half the de Broglie wavelength \cite{Lee84}$^{:22}$) for us, since the latter
is only applicable for $T<\allowbreak0.92$ eV.

$\allowbreak\gamma_{\text{U1}}^{\text{e}}\left(  2\right)  $ is a function of
(Coulomb logarithm) $\mathcal{L}$ that significantly improves the accuracy
down to about $\mathcal{L}=2.5$ \cite{Viegas71a}. The minimum $\mathcal{L}$
possible for our results is $2.8$ at the aforementioned extremes, although
$\mathcal{L}$ only goes down to $3.56$ for the S-band data analyzed. Note that
we use Viegas' definition for $\mathcal{L}$ since that is the one that
$\gamma_{\text{U1}}^{\text{e}}\left(  2\right)  $ assumes. The derivation of
$\gamma_{\text{U1}}^{\text{e}}\left(  2\right)  $ entails use of an integral
upon which the first choices for $b_{\max}$ and $b_{\min}$ are based
\cite{Viegas71b}$^{:9}$\cite{Itikawa63}$^{:2.8}$. The error in $\gamma
_{\text{U1}}^{\text{e}}\left(  2\right)  $ for cases where a second choice is
used is not known, but assumed to be less that if it was not used at all.
Fortunately, this only affects the most extreme cases for our application.

The momentum transfer cross section from which $\sigma_{\text{SH}}$ is derived
\cite{Goldston95}$^{:11.19}$\cite{Bittencourt04}$^{:20.8.13}$%
\cite{Gilardini72}$^{:2.6.25}$ has a $v^{-4}$ dependence. From this, the
solution to $q_{3}\left(  v\right)  $ in Eqs.~\ref{X13} for $\nu_{\text{c}3}$
and $\omega=0$ that results in Eq.~\ref{X14}.1 is,
\begin{equation}
q_{3}\left(  v\right)  =\frac{\mathcal{L}}{\gamma_{E}\gamma_{\text{U1}%
}^{\text{e}}\left(  2\right)  }\frac{e^{4}}{4\pi\epsilon_{0}^{2}m^{2}v^{4}}
\label{X15}%
\end{equation}

Figure~10 plots $\sigma\left(  0\right)  $ vs.\ $I_{0}$ (Eq.~\ref{X13}.1) for
the our values of $p$. The results are compared to those assuming $\omega=0$,
showing it to be a small effect for all but $p=1.11$~Torr. Aside from this,
the significant differences in the $\sigma$ curves for $I_{0}\gtrsim10^{14}%
$~W/cm$^{2}$ are due to the dependence of $\mathcal{L}$ on electron density
for $\nu_{3}$. Electron-ion collisions dominate those of neutrals in this
range, as indicated by $\nu_{\text{c}3}>\nu_{\text{c}1}+\nu_{\text{c}2}$ for
$I_{0}\gtrsim1.0\times10^{14}$~W/cm$^{2}$ beyond the value of $v$ for which
the exponential term in Eq.~\ref{X13}.2 is equal to $e^{-1}$. Figure~11 uses
the results of this section to calculate values of properties of interest, as
determined by the $\sigma\left(  z\right)  $ measurements.%

\begin{figure}[H]\includegraphics[width=3.375in]{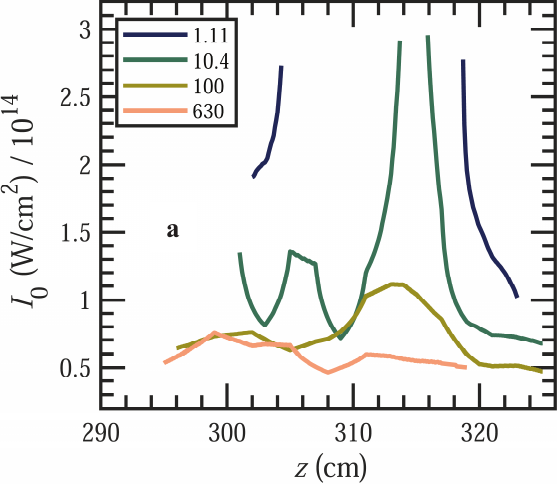}\end{figure}           
\begin{figure}[H]\includegraphics[width=3.375in]{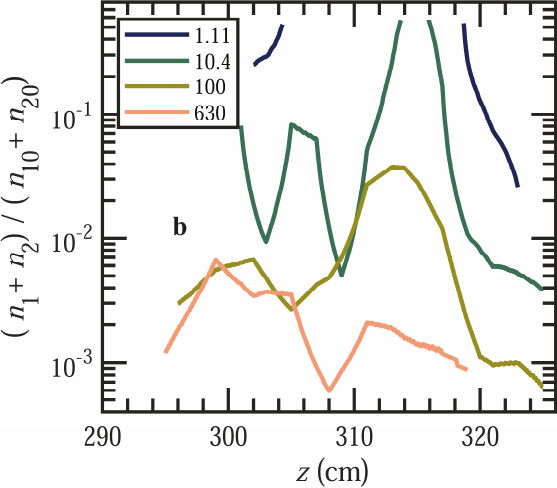}\end{figure}           
\begin{figure}[H]\includegraphics[width=3.375in]{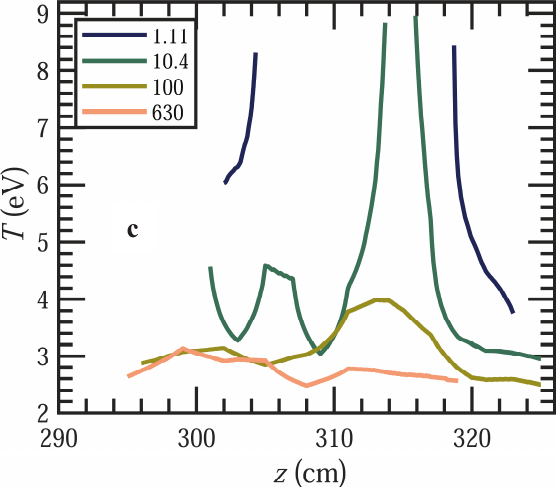}\end{figure}           

\textbf{Fig.~11} Based on our $\sigma\left(  z\right)  $ measurements, plot
(a) is $I_{0}\left(  T\left(  \sigma\left(  z\right)  \right)  \right)  $,
where $I_{0}\left(  T\right)  $ is the inverse of $T\left(  I_{0}\right)  $
($T_{2}$ in reference \cite{Ruden25QmodT}) and $T\left(  \sigma\right)  $ is
the inverse of $\sigma\left(  n_{1}\left(  I_{0}\left(  T\right)  \right)
,n_{2}\left(  I_{0}\left(  T\right)  \right)  ,T\right)  $, where
$\sigma\left(  n_{1},n_{2},T\right)  $ is the solution to Eq.~\ref{X13}.1 with
$\omega=\omega_{10}$, and $n_{1}\left(  I_{0}\right)  $ and $n_{2}\left(
I_{0}\right)  $ are solutions to Eqs.~\ref{X12}. Plot (b) is $n_{1}\left(
I_{0}\left(  T\left(  \sigma\left(  z\right)  \right)  \right)  \right)  $ and
$n_{2}\left(  I_{0}\left(  T\left(  \sigma\left(  z\right)  \right)  \right)
\right)  $. Plot (c) is\textbf{ }$T\left(  \sigma\left(  z\right)  \right)  $.
Background pressure $p$ is specified in Torr in the legends. The upward surges
for lower $p$ values are due to excessive violation of the single-filament
approximation and/or parameters exceeding values modeled.

\bigskip

Although is S-band waveguide is too slow to measure the filament current time
history, the current time integral $Q$ (insensitive to the details of the
current waveform) is determined from the self-emission signal. $V_{\text{R}}$
due to self-emission recorded directly from the $50$~$\Omega$ adapter port
(with no HP\ crystal detector) was not routinely recorded due to low signal
levels. $V_{\text{R}}$ increases significantly as $p$ drops, though, and a
record suitable for comparison with the simulations was obtained at
$p=50$~Torr near the center of the filament ($z=307$~cm) by averaging over
$200$ shots. $V_{\text{R}}$ after $3$~ns, as plotted in Fig.~12 (lower trace),
appears rather chaotic, but is very reproducible after such averaging.
$V_{\text{R}}$ varies little in shape, regardless of $p$ and $z$, changing
primarily only in amplitude, as implied by Fig.~2.

A number of self-emission simulations with various $I_{0}\left(  t\right)  $
waveforms, filament radii, and trailing conductivities were performed to
calibrate the signal. Details are provided in Appx.~A. The best match, also
shown in Fig.~12 (upper trace), is obtained with a simulated filament with
$R=25.1$~$\mu$m (base 10 $\log\left(  R\text{ }\left(  \mu\text{m}\right)
\right)  =1.4$) and a post-current conductivity of $\sigma=5.09\times10^{5}$
S/m ($G=10^{-4}$ Sm). This, we will see, is a small radius relative to that
measured, but with a comparable value of $G$. The simulation uses a negative
going Gaussian $I_{0}\left(  t\right)  $ with a (very short) $5$ ps standard
deviation width and a total time integral of $Q_{\text{sim}}=-1$ pC. The
intention is to approximate the effect of a point particle with this charge.
Based on the relative peak to peak amplitude of the highest amplitude bipolar
oscillation of the experimental vs.\ simulated $V_{\text{R}}$ waveforms
plotted in Fig.~12, the experimental value of $Q$ is $Q_{50}=0.93$~pC.

\section{Charge transfer $Q$ measured by self-emission}%

\begin{figure}[H]\includegraphics[width=3.375in]{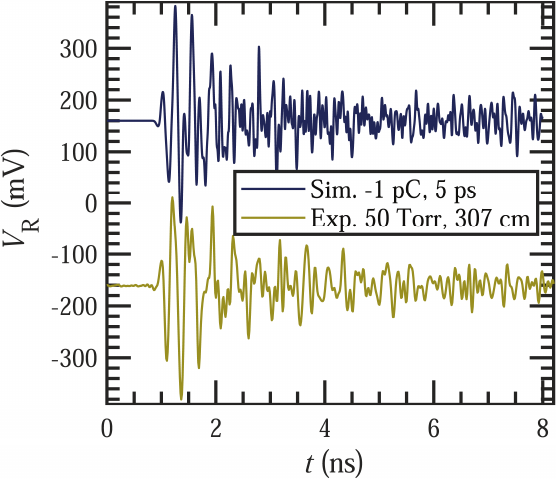}\end{figure}           

\textbf{Fig.~12 }An experimentally measured $V_{\text{R}}$ signal
$V_{\text{R50}}$ (bottom) recorded directly from the waveguide receiver output
into a $50$~$\Omega$ port with the waveguide centered at $z=307$~cm due to
self-emission from a filament formed at $p=50$~Torr is compared here to a
simulated $V_{\text{R}}$ self-emission signal $V_{\text{Rsim}}$ resulting from
a Gaussian current waveform with $5$ ps standard deviation and an
$I_{0}\left(  t\right)  $ time integral of $Q=-1$~pC (top). The traces are
vertically offset from their $V_{\text{R}}=0$~mV initial baseline to avoid overlap.

\bigskip

Neglecting energy reabsorbed and dissipated by the filament itself, the total
detector output energy measured by the power detector (time integrated
$P_{\text{R}}$ from Eq.~\ref{X1}) due to self-emission is theoretically
proportional to the square of $Q$ due to the EM field intensity scaling
linearly with $Q$. We, therefore, have $Q$ vs.\ $z$ and $p$,%
\begin{equation}
Q=Q_{50}\sqrt{\frac{\int_{0}^{\infty}P_{\text{R}}dt\text{ }}{\int_{0}^{\infty
}P_{\text{R50}}dt}}\ \label{X16}%
\end{equation}
where $P_{\text{R50}}$ is $P_{\text{R}}$ measured \emph{with the detector
attached} under the same test conditions as the direct $V_{\text{R50}}$
measurement ($p=50$~Torr, $z=307$~cm).

Note that the accuracy of $K$\ affects the absolute accuracy of $P_{\text{R}}%
$, based on the $V_{\text{D}}$ measurement (Eq.~\ref{X1}), but not $Q$ itself,
since $K$ cancels out in the ratio. The correlation of experimental
$V_{\text{R}}$ is very good up to $t_{1}=1.4$ ns, so this value is used for
the integration limit. The subsequent simulated signal likely has
contributions from the high frequency artifacts resulting from the filament
axial mesh. Figure~13 plots the results. The continuous lines between actual
data (symbols) is the sinusoidal expansion discussed in Sec.~VII.%

\begin{figure}[H]\includegraphics[width=3.375in]{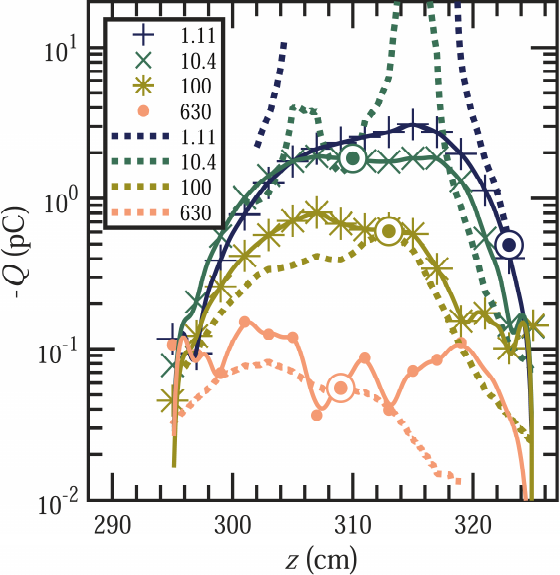}\end{figure}           

\textbf{Fig.~13} The legend symbols are $-Q$ (negative time integrated
current) measurements for a range of background pressures $p$, based on the
square root of the crystal detector signal time integral\ (Eqs.~\ref{X16}) as
a function of distance $z$ of the waveguide center from the spherical mirror
used to focus the USPL laser pulse. The connecting solid lines are the
sinusoidal expansions of the piecewise linear function connecting these data
points out to $32$ modes (Eqs.~\ref{X34}). The dotted lines plot $-Q$, based
on Eq.~\ref{X23}, Eq.~\ref{X26}, and $Q=I_{\text{f}}/\nu_{1}$. $p$ is
specified in Torr in the legend. The circled dots identify that locations
$z=z_{1}$ (Table~2) nearest the central $z=z_{0}=310$~cm where the two $Q$
estimates are equal (Table~2).

\section{Current decay rate $\nu$ and $Q$ for a steady state filament}

The temperature model \cite{Ruden25QmodT} used above also provides the
full-pulse average electron axial momentum $\left\langle p_{{z}}\right\rangle
$ ($\left\langle p_{2{\mathrm{f}z}}\right\rangle _{{\gamma}}$ in this
reference) vs.\ laser intensity\ $I_{0}$ resulting from the optical
$\mathbf{J\times B}$ force on an electron that rescatters off its parent ion
within an optical cycle after release. Along with $Q$, the rate $\nu$ at which
the time-dependant filament current $I_{\text{f}t}\equiv I_{\text{f}t}\left(
t\right)  $ decays due to Ohmic dissipation plays an important role in far
field EM radiation in the $\sim f_{1}$ band of interest for sufficiently large
$\theta$ (to be quantified). In this section, we neglect the $z$\ dependence
of $Q$, and assume a steady state filament uniform in $r$ with radius
$R=1.9243s$ (Eq.~\ref{X9}), for our $\nu$ estimate.

The post-optical filament current and ion species' densities resulting from
the laser interaction exclusively are approximated as, \ \ \ %

\begin{equation}%
\begin{tabular}
[c]{l}%
$I_{\text{f}0}=-\frac{e\pi R^{2}n_{R}}{m}\left\langle p_{{z}}\right\rangle $,
\ $n_{R}=n_{R1}+n_{R2}$\\
$n_{Ri}=\frac{2s^{2}}{R^{2}}n_{i}=0.5401n_{i}$, \ $i=1,2$%
\end{tabular}
\ \ \ \ \label{X17}%
\end{equation}
Here, we have assumed the same Gaussian profile for $n_{i}$, as we did for
$\sigma\left(  r\right)  $ and $J\left(  r\right)  $ to determine the
corresponding uniform $n_{Ri}$ (analogous to Eq.~\ref{X6}). From this, the
post-optical center-of-mass electron kinetic energy per unit length imparted
by the laser is expressed as $U_{0}=\lambda_{I}I_{\text{f}0}^{2}$, where
$\lambda_{I}$ is defined in Eqs.~\ref{X25} below.

Collective effects cause some of $U_{0}$ to be converted rapidly (relative to
our $f_{1}^{-1}$ time scale) into near field EM energy. Radiated energy is
neglected in this section since the steady state assumption describes a rigid
charge distribution moving at constant velocity, which does not radiate. In
terms of the reduced current $I_{\text{f}}$ after such conversion, the
time-dependent current is approximated by $I_{\text{f}t}=I_{\text{f}}%
\exp\left(  -\nu t\right)  $ (oscillating modes are negected). $t=0$ is the
time of optical pulse passage in this section.

One estimate of $\nu=\nu_{1}$ is found by assuming the near field energy is
the minimum consistent with Maxwell's equations and the following electron
momentum equation, based on a generalized form of Ohm's law \cite{Kimura14},%
\begin{equation}
\mathbf{E}=\frac{\mathbf{J}}{\sigma}+\frac{m}{e^{2}n_{R}}\frac{\partial
\mathbf{J}}{\partial t}+\frac{1}{en}\left(  \mathbf{J}\times\mathbf{B}-\nabla
p_{e}\right)  \label{X19}%
\end{equation}
where $p_{e}$ is electron pressure. The third (Hall and pressure gradient)
term on the r.h.s.\ is neglected, though. From this and Eq.~\ref{X13} then,
\ \ \ \ \
\begin{equation}
\frac{e^{2}n_{R}}{m}\mathbf{E}=\left(  \nu_{\text{c,eff}}+\frac{\partial
}{\partial t}\right)  \mathbf{J} \label{X20}%
\end{equation}
This is the time domain representation of the relationship between
$\mathbf{E}$ and $\mathbf{J}$ assumed by the\ Drude model \cite{Adamyan09} of
frequency-dependent conductivity. That is, the time domain basis of Drude
conductivity is no longer neglected. In the absence of an external driver,
Eqs.~\ref{X13} with $\omega=0$ and $n_{i}=n_{Ri}$ of Eq.~\ref{X17} are used
for $\nu_{\text{c,eff}}$.

Only the (dominant) $z$ component of $\mathbf{E}$ and $\mathbf{J}$ and the
cylindrical coordinate azimuthal $\varphi$ component of $\mathbf{B}$ are
considered. This simplifies the analysis, but neglects higher order components
that result from the neglected third term of Eq.~\ref{X19} as the filament
acquires a net charge $Q$ trailing the pulse. Taking the curl of Ampere's law;
replacing the resulting $\nabla\times\nabla\times\mathbf{B}$ term with its
identity $\nabla\left(  \nabla\cdot\mathbf{B}\right)  \mathbf{-}\nabla
^{2}\mathbf{B}$, where $\nabla\cdot\mathbf{B=0}$; operating on the result with
$\left(  \nu_{\text{c,eff}}+\partial/\partial t\right)  $; distributing and
commuting this operator to isolate $\left(  \nu_{\text{c,eff}}+\partial
/\partial t\right)  \mathbf{J}$; replacing that with $n_{R}e^{2}\mathbf{E}/m$
based on Eq.~\ref{X20}; and using Faraday's law to replace the resultant two
occurrences of $\nabla\times\mathbf{\mathbf{E}}$ with $-\partial
\mathbf{B/}\partial t$ results in,%
\begin{equation}
\frac{\mu_{0}e^{2}n_{R}}{m}\frac{\partial\mathbf{B}}{\partial t}=\left(
\nu_{\text{c,eff}}+\frac{\partial}{\partial t}\right)  \left(  \nabla
^{2}\mathbf{B-}\epsilon_{0}\mu_{0}\frac{\partial^{2}\mathbf{B}}{\partial
t^{2}}\right)  \label{X21}%
\end{equation}

Equation~\ref{X21} has a number of solutions (modes). We seek decaying
solutions with a $\varphi$ component only of the form $B_{j\varphi}e^{-\nu
_{j}t}$, where $\nu_{j}$ is a positive real number. Taking $\partial/\partial
t\rightarrow-\nu_{j}$ and $\partial/\partial z\rightarrow-c^{-1}%
\partial/\partial t=c^{-1}\nu_{j}$ for a steady state solution, Eq.~\ref{X21}
becomes,%
\begin{equation}%
\begin{tabular}
[c]{l}%
$s_{j}^{2}\frac{\partial^{2}B_{j\varphi}}{\partial s_{j}^{2}}+s_{j}%
\frac{\partial B_{j\varphi}}{\partial s_{j}}+\left(  s_{j}^{2}-1\right)
B_{j\varphi}=0$\\
$s_{j}=r\sqrt{\frac{\mu_{0}e^{2}n_{R}\nu_{j}}{m\left(  \nu_{\text{c,eff}}%
-\nu_{j}\right)  }+\left[  c^{-2}\mathbf{-}\epsilon_{0}\mu_{0}\right]  \nu
_{j}^{2}}=r\sqrt{\frac{\mu_{0}e^{2}n\nu_{j}}{m\left(  \nu_{\text{c,eff}}%
-\nu_{j}\right)  }}$%
\end{tabular}
\ \ \label{X22}%
\end{equation}

Equation \ref{X22}.1 is the Bessel equation of order $1$ for uniform $n$, with
solutions $B_{j\varphi}=$ $b_{j}J_{1}\left(  s_{j}\right)  $ for $r\leq R$,
where $b_{j}$ is a constant and $J_{1}\left(  s\right)  $ is the Bessel
function of the first kind of order $1$. The $c^{-2}$ in square brackets
results from a $\partial^{2}/\partial z^{2}\rightarrow c^{-2}\nu_{j}^{2}$
substitution and, therefore, refers to the speed $c$ of the USPL pulse.
$\epsilon_{0}\mu_{0}$ subtracted from it, though, refers to EM properties of
neutral air at microwave frequencies. These approximately equal their vacuum
values and cancel in cases of interest for the third term. Given this,
$B_{j\varphi}=0$ for $r>R$ since $s_{j}=n=0$ there and $J_{1}\left(  0\right)
=0$. $B_{j\varphi}$ is continuous across the boundary, so the interior
solution has $s_{j}=S_{j}$ at $r=R$, where $S_{j}$ is the value of $s$ at the
$j$'th zero crossing of $J_{1}\left(  s\right)  $. Solving Eq.~\ref{X22}.2 for
$\nu_{j}$ at $r=R$,
\begin{equation}
\nu_{j}=\frac{mS_{j}^{2}\nu_{\text{c,eff}}}{\mu_{0}e^{2}R^{2}n_{R}+mS_{j}^{2}}
\label{X23}%
\end{equation}
As before, we have neglected the imaginary contribution to collision frequency
$\nu_{\text{c,eff}}$ in Eq.~\ref{X13}.2. Such a term contributes to
oscillatory Sommerfeld wire surface modes \cite{Stratton41}$^{:6.4.6}$ that
have little effect on total charge transfer due to the monotonically decaying
current of interest here. Such modes are left to a more detailed nonlinear
numerical treatment \cite{Garrett21}\cite{Garrett25}.

$I_{\text{f}}$ is found by assuming that center-of-mass energy loss is limited
to that of the EM field of the primary ($j=1$) mode of Eq.~\ref{X22}, and that
only components $J_{z}$, $E_{z}$, and the $B_{\varphi}$ solution to
Eq.~\ref{X22}.1 are considered. Faraday's law then implies $\partial
E_{jz}/\partial r-\nu_{j}E_{jr}/c=\nu_{j}B_{j\varphi}$, given $\partial
/\partial z\rightarrow c^{-1}\nu_{j}$ and $\partial/\partial t\rightarrow
-\nu_{j}$, where $\mathbf{E}$ solutions are of the form $\mathbf{E}_{j}%
\exp\left(  -\nu_{j}t\right)  $. Neglecting $E_{jr}$, this expression implies
$E_{jz}$ is constant for $r\geq R$, since $B_{j\varphi}=0$ there. And, since
$E_{jz}=0$ as $r\rightarrow\infty$, we have $E_{jz}=B_{jz}=0$ for $r\geq R$.
Our solution to $B_{j\varphi}$, the identity $dJ_{0}\left(  s\right)
/ds=\allowbreak-J_{1}\left(  s\right)  $, and the relationship between
$\mathbf{J}$ and $\mathbf{E}$ in Eq.~\ref{X20} after pulse passage then give
us, for $r\leq R$,
\begin{equation}%
\begin{tabular}
[c]{l}%
$B_{\varphi}=\sum_{j=1}^{\infty}b_{j}J_{1}\left(  s_{j}\right)  e^{-\nu_{j}t}%
$\\
$E_{z}=-\sum_{j=1}^{\infty}b_{j}\sqrt{\frac{m\left(  \nu_{\text{c,eff}}%
-\nu_{j}\right)  \nu_{j}}{\mu_{0}e^{2}n_{R}}}\left(  J_{0}\left(
s_{j}\right)  -J_{0}\left(  S_{j}\right)  \right)  e^{-\nu_{j}t}$\\
$J_{z}=-\sum_{j=1}^{\infty}b_{j}\sqrt{\frac{n_{R}e^{2}\nu_{j}}{\mu_{0}m\left(
\nu_{\text{c,eff}}-\nu_{j}\right)  }}\left(  J_{0}\left(  s_{j}\right)
-J_{0}\left(  S_{j}\right)  \right)  e^{-\nu_{j}t}$%
\end{tabular}
\ \ \ \label{X24}%
\end{equation}

The post-optical energies per unit axial length due to $E_{z}$ and
$B_{\varphi}$ are found by integrating $2\pi r\epsilon_{0}E_{z}^{2}/2$ and
$2\pi rB_{\varphi}^{2}/\left(  2\mu_{0}\right)  $, respectively, from $r=0$ to
$R$ at $t=0$. Equating $I_{\text{f}}$ to the integral of $2\pi rJ_{z}$ over
$r$ out to $R$ for $j=1$ (only) allows us to eliminate $b_{1}^{2}$ in favor of
$I_{\text{f}}^{2}$ in Eqs.~\ref{X24}. This permits the energies per unit
length to be expressed as $\lambda_{E}I_{\text{f}}^{2}$ and $\lambda
_{B}I_{\text{f}}^{2}$, respectively. From the above,%

\begin{equation}%
\begin{tabular}
[c]{l}%
$\lambda_{I}=\frac{m}{2\pi e^{2}R^{2}n_{R}}$ $\ \ \ \ \ \ \ \ \ \ \lambda
_{B}=\frac{a_{1}\mu_{0}}{4\pi S_{1}^{2}J_{0}^{2}\left(  S_{1}\right)  }$\\
$\lambda_{E}=\frac{\epsilon_{0}\left(  a_{0}+S_{1}J_{0}^{2}\left(
S_{1}\right)  \right)  \left(  \mu_{0}mR\right)  ^{2}\nu_{\text{c,eff}}}{4\pi
J_{0}^{2}\left(  S_{1}\right)  \left(  \mu_{0}e^{2}R^{2}n_{R}+mS_{1}%
^{2}\right)  ^{2}}$ \ \ $S_{1}=3.8317$\\
$\int_{0}^{S_{1}}s_{1}J_{0}\left(  s_{1}\right)  ds_{1}=\allowbreak S_{1}%
J_{1}\left(  S_{1}\right)  =0$\\
$a_{0}=\int_{0}^{S_{1}}s_{1}J_{0}^{2}\left(  s_{1}\right)  ds_{1}=$ $_{1}%
F_{2}\left(  \frac{1}{2};1,2;-S_{1}^{2}\right)  \frac{S_{1}^{2}}{2}$\\
$a_{1}=\int_{0}^{S_{1}}s_{1}J_{1}^{2}\left(  s_{1}\right)  ds_{1}\allowbreak$
$=$ $_{1}F_{2}\left(  \frac{3}{2};3,3;-S_{1}^{2}\right)  \frac{S_{1}^{2}}{16}%
$\\
$a_{0}=a_{1}=\allowbreak1.190\,8$ \ \ \ $J_{0}\left(  S_{1}\right)
=-0.402\,76$%
\end{tabular}
\ \ \ \label{X25}%
\end{equation}
where Eq.~\ref{X23} has been used to eliminate $\nu_{1}$ in favor of
$\nu_{\text{c,eff}}$ for the $\lambda_{E}$ solution. We have included here the
identities used to solve for $\lambda_{E}$ and $\lambda_{B}$. $_{1}F_{2}$
refers to the generalized hypergeometric series \cite{Gradshteyn15}%
$^{:9.14.1}$. $\nu_{\text{c,eff}}$ is found from Eq.~\ref{X13}.2. However, the
current contribution considered is unipolar and no longer driven by an
external $\mathbf{E}$, so we set $\omega=0$. Also, $n_{Ri}$ from
Eq.~\ref{X17}.2 replaces $n_{i}$ in Eq.~\ref{X13}.3, Eq.~\ref{X13}.4, and
Eq.~\ref{X14}.2.%

\begin{figure}[H]\includegraphics[width=3.375in]{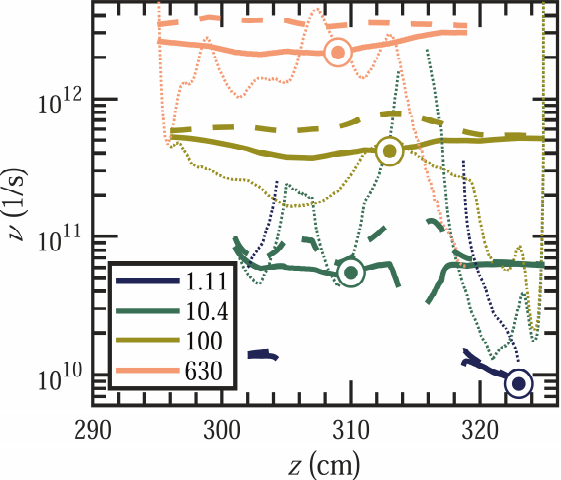}\end{figure}           

\textbf{Fig.~14} Effective electron collision frequency $\nu_{\text{c,eff}}$
(Eqs.~\ref{X13} with $\omega=0$ and $n_{i}=n_{Ri}$ of Eq.~\ref{X17}, dashed
lines), bulk current decay rate $\nu_{1}$ (Eq.~\ref{X23}, solid lines), and
surface current decay rate $\nu_{0}$ (Eqs.~\ref{X27}, thin dotted lines), both
assuming a steady state filament (with $\partial Q\left(  z\right)  /\partial
z=0$), are plotted. Background pressure $p$ is specified in Torr in the
legend. The circled dots identify where $\nu=\nu_{1}$ at $z=z_{1}$ (Table~2)
used for the constant $\nu$ values needed for the far field convolution of
Eqs.~\ref{X37}. Constant $\nu_{0}$ values for the convolution are taken from
the same $z$ locations.

\bigskip

Neglecting other forms of energy conversion, we have%
\begin{equation}
I_{\text{f}}=I_{\text{f}0}\sqrt{\frac{\lambda_{I}}{\lambda_{I}+\lambda
_{\ E}+\lambda_{B}}} \label{X26}%
\end{equation}
We find for our problem that $\lambda_{B}$ is generally comparable to
$\lambda_{I}$, while $\lambda_{E}$ is negligible. Figure~14 plots the
corresponding values of $\nu_{1}$ and $\nu_{\text{c,eff}}$. Although
$\nu_{\text{c,eff}}$ is generally no more that twice $\nu_{1}$, a significant
degree of thermalization can still be assumed since the model used to
determine $T\left(  I_{0}\right)  $ \cite{Ruden25QmodT} takes into account an
extra prompt optical intracycle rescatter of the electron off its parent ion.
As a bonus, $Q=I_{\text{f}}/\nu_{1}$ is an independent estimate of $Q$ (albeit
for a steady state filament) for comparison to that directly measured, as
plotted in Fig.~13.

The EM field energized by $U_{0}$ in the above is confined to the filament and
distributed through the bulk like the electrons accelerated by the laser
themselves. A complementary estimate of $\nu=\nu_{0}$ suggested buy 3-D time
domain simulations \cite{Garrett21}\cite{Garrett25} is found by assuming the
current and, therefore, the near\ field it produces, is expelled to the
periphery of the filament during thermalization. Neglecting oscillatory modes,
as before, Ohmic dissipation of $U_{0}$ is assumed to be due to an
exponentially decaying current that diffuses back in. We make direct use of
the direct self-emission $Q$ measurement for this, and (again) assume that the
near field energy is exponentially dissipated Ohmically.

With $I_{\text{f}t}=I_{\text{f}}\exp\left(  -\nu_{0}t\right)  =v_{0}%
Q\exp\left(  -\nu_{0}t\right)  $, then, we equate the laser-imparted
$U_{0}=\lambda_{I}I_{\text{f}0}^{2}$ (from Eq.~\ref{X17}.1 and Eq.
\ref{X25}.1) to the total time integral of the resistive heating rate per unit
length $I_{\text{f}t}^{2}\mathcal{R}$, where $\mathcal{R=}\pi\sigma\left(
2R-D\right)  D$ is the resistance per unit length (thick-walled tube
approximation \cite{Wheeler42}) and $D$ is the skin depth for a exponentially
decaying current. This implies,%
\begin{equation}
\nu_{0}=\frac{2\pi U_{0}\sigma\left(  2R-D\right)  D}{Q^{2}}\text{
\ \ \ \ }D=\sqrt{\frac{1}{\mu_{0}\sigma\nu_{0}}}\leq R \label{X27}%
\end{equation}
$D$, as expressed here, is found by assuming an $e^{-\nu_{0}t}$ variation in
its derivation \cite{Jackson99}$^{:8.5-8.15}$ instead of $e^{-i\omega t}$ for
real $\nu_{0}$ and $\omega$. We base $\sigma$ on the same $\nu_{\text{c,eff}}$
used for $\nu_{1}$ in Eq.~\ref{X13}.1, with $n_{1}+n_{2}=n_{R}$ from
Eq.~\ref{X17}. $\nu_{0}$ is found by substituting the right expression into
the left, and expressing the result as a quartic in $\nu_{0}^{-1/2}$. Solved
numerically, a physically meaningful root is identified as being real and
resulting in $D\leq R$. Not having one implies strong diffusion, so $D=R$ is
set. $D$ so determined is then substituted into the left expression to
determine $\nu_{0}$ (Fig.~14). Its much greater volatility vs. $\nu_{1}$ is
due to its dependence on the results plotted in Fig.~11. $\nu_{0}$ at the
$z=z_{1}$ reference points are comparable to $\nu_{1}$ (Table~2) due to
significant diffusion ($D>0.4R$) at $z=z_{1}$ for all $p$.

\section{Far field radiation from $Q=Q\left(  z\right)  $ and $\nu$}

The above calculations neglect energy loss to the far field that results from
changes in $Q$. We now estimate the far field due to the $z$-dependence of
$Q=$ $Q\left(  z\right)  $, assuming a constant $\nu$ for each given $p$ at
$z=z_{1}$. For an impulsive current, though, the current and charge densities
are approximated by,%

\begin{equation}%
\begin{tabular}
[c]{c}%
$\mathbf{J}_{0}\left(  \mathbf{x},t\right)  =cQ\left(  z\right)  \delta\left(
x\right)  \delta\left(  y\right)  \delta\left(  z-z_{0}-ct\right)
\mathbf{\hat{z}}$ \ \ \ \ \\
$\rho_{0}\left(  \mathbf{x},t\right)  =Q\left(  z\right)  \delta\left(
x\right)  \delta\left(  y\right)  \delta\left(  z-z_{0}-ct\right)  $\\
$-Q^{\prime}\left(  z\right)  \delta\left(  x\right)  \delta\left(  y\right)
\Theta\left(  ct-z+z_{0}\right)  $%
\end{tabular}
\ \label{X28}%
\end{equation}
where $\delta\left(  z\right)  $ is the Dirac delta function, $\Theta\left(
z\right)  $ is the Heaviside step function. Zero subscripts here and below are
used to identify impulse response functions. The $Q^{\prime}\left(  z\right)
\equiv\partial Q\left(  z\right)  /\partial z$ term is the net charge density
left behind the pulse implied by charge conservation. Time $t=0$ here is now
taken to be when the pulse is at $z=z_{0}=310$~cm (deemed the center of the filament).

The magnetic vector potential $\mathbf{A}_{0}$ and electric scalar potential
$\Phi_{0}$ are determined from Eq.~\ref{X28} by integration of the retarded
contributions over space \cite{Jackson99}$^{:6.48}$,%
\begin{equation}%
\begin{tabular}
[c]{c}%
$\mathbf{A}_{0}=\frac{\mu_{0}}{4\pi}%
{\displaystyle\int_{-\infty}^{+\infty}}
\frac{cQ\left(  z^{\prime}\right)  \delta\left(  z^{\prime}-z_{0}+\left\vert
\mathbf{x}-\mathbf{x}^{\prime}\right\vert -ct\right)  }{\left\vert
\mathbf{x}-\mathbf{x}^{\prime}\right\vert }dz^{\prime}\mathbf{\hat{z}}$\\
$\Phi_{0}=\frac{1}{4\pi\epsilon_{0}}%
{\displaystyle\int_{-\infty}^{+\infty}}
\frac{Q\left(  z^{\prime}\right)  \delta\left(  z^{\prime}-z_{0}+\left\vert
\mathbf{x}-\mathbf{x}^{\prime}\right\vert -ct\right)  }{\left\vert
\mathbf{x}-\mathbf{x}^{\prime}\right\vert }dz^{\prime}$\\
$\ \ \ \ \ \ -\frac{1}{4\pi\epsilon_{0}}%
{\displaystyle\int_{-\infty}^{+\infty}}
\frac{Q^{\prime}\left(  z^{\prime}\right)  \Theta\left(  ct-\left\vert
\mathbf{x}-\mathbf{x}^{\prime}\right\vert -z^{\prime}+z_{0}\right)
}{\left\vert \mathbf{x}-\mathbf{x}^{\prime}\right\vert }dz^{\prime}$\\
$\left\vert \mathbf{x}-\mathbf{x}^{\prime}\right\vert =\sqrt{r^{2}\sin
^{2}\theta+\left(  r\cos\theta-z^{\prime}+z_{0}\right)  ^{2}}$\\
$\mathbf{\hat{z}=\hat{r}}\cos\theta-\mathbf{\hat{\theta}}\sin\theta$%
\end{tabular}
\ \ \label{X29}%
\end{equation}
The last two lines result in $\mathbf{A}_{0}$ and $\Phi_{0}$ expressed in
terms of $r$ and $\theta$ , where $r$ (now) refers to the spherical coordinate
radius centered at $x=y=0$ and $z=z_{0}$ and $\theta$\ is the angle relative
to the $z$-axis. Changing the integration variable to $\tau=\left(  z^{\prime
}-z_{0}+\left\vert \mathbf{x}-\mathbf{x}^{\prime}\right\vert \right)  /c$ we
have,
\begin{equation}%
\begin{tabular}
[c]{l}%
$c\tau\equiv z^{\prime}+\sqrt{r^{2}\sin^{2}\theta+\left(  r\cos\theta
-z^{\prime}+z_{0}\right)  ^{2}}$\\
$z^{\prime}-z_{0}=\frac{\left(  c\tau{}-r\right)  \left(  c\tau{}+r\right)
}{2\left(  c\tau-r\cos\theta\right)  }$ \ \ \ $dz^{\prime}=\frac{c\left(
c^{2}\tau-2c\tau r\cos\theta+r^{2}\right)  }{2\left(  c\tau-r\cos
\theta\right)  ^{2}}d\tau$%
\end{tabular}
\ \ \label{X30}%
\end{equation}
This results in,
\begin{equation}%
\begin{tabular}
[c]{c}%
$\mathbf{A}_{0}=\frac{\mu_{0}c}{4\pi}\frac{Q\left(  \frac{\left(
ct{}-r\right)  \left(  ct{}+r\right)  }{2\left(  ct-r\cos\theta\right)
}\right)  }{\left(  ct-r\cos\theta\right)  }\left(  \mathbf{\hat{r}}\cos
\theta-\mathbf{\hat{\theta}}\sin\theta\right)  $\\
$\Phi_{0}=\frac{1}{4\pi\epsilon_{0}}\frac{Q\left(  \frac{\left(  ct-r\right)
\left(  ct+r\right)  }{2\left(  ct-r\cos\theta\right)  }\right)  }{\left(
ct-r\cos\theta\right)  }$\\
$-\frac{1}{4\pi\epsilon_{0}}%
{\displaystyle\int_{-\infty}^{t}}
\frac{cQ^{\prime}\left(  \frac{\left(  c\tau-r\right)  \left(  ct{}^{\prime
}+r\right)  }{2\left(  c\tau-r\cos\theta\right)  }\right)  }{\left(
c\tau-r\cos\theta\right)  }d\tau\ \ \ \ \ \ $%
\end{tabular}
\ \ \label{X31}%
\end{equation}

We replace $\left(  ct+r\right)  $ with $2r$ and $\left(  ct-r\cos
\theta\right)  $ with $r\left(  1-\cos\theta\right)  $ in Eqs.~\ref{X31} for
the far field since $ct\approx r$ there. The integrand of the second $\Phi$
term then reduces to$\ \partial Q\left(  \frac{c\tau-r}{1-\cos\theta}\right)
/\partial\tau$, which integrates to $Q\left(  \frac{ct-r}{1-\cos\theta
}\right)  $ and adds to the first $\Phi_{0}$ term. The far field is then,%

\begin{equation}%
\begin{tabular}
[c]{l}%
$\mathbf{A}_{0}=\frac{\mu_{0}c}{4\pi}\frac{Q\left(  \frac{ct-r}{1-\cos\theta
}\right)  }{r\left(  1-\cos\theta\right)  }\left(  \mathbf{\hat{r}}\cos
\theta-\mathbf{\hat{\theta}}\sin\theta\right)  $\\
$\Phi_{0}=\frac{1}{4\pi\epsilon_{0}}\frac{Q\left(  \frac{ct-r}{1-\cos\theta
}\right)  \cos\theta}{r\left(  1-\cos\theta\right)  }$%
\end{tabular}
\ \label{X32}%
\end{equation}
From these, $\mathbf{E}_{0}=-\nabla\Phi_{0}-\partial\mathbf{A_{0}/}\partial t$
\cite{Jackson99}$^{:6.9}$. The $r$ component of $\nabla\Phi_{0}$ cancels that
of $\partial\mathbf{A_{0}/}\partial t$, leaving only the $\theta$ components
of each. Contributions to $\nabla\Phi_{0}$ that fall off as $1/r^{2}$ are
discarded as not contributing to the far field. We have, then,%

\begin{equation}
\mathbf{E}_{0}\left(  t\right)  =\frac{Q^{\prime}\left(  \frac{ct-r}%
{1-\cos\theta}\right)  \left(  \sin\theta-\sin2\theta\right)  }{4\pi
\epsilon_{0}r\left(  1-\cos\theta\right)  ^{3}}\mathbf{\hat{\theta}}
\label{X33}%
\end{equation}
$E_{\theta}$ decreases rapidly with increasing $\theta$ until $\theta
=60^{\circ}$, where it crosses zero, and doesn't rise significantly beyond that.%

\begin{figure}[H]\includegraphics[width=3.375in]{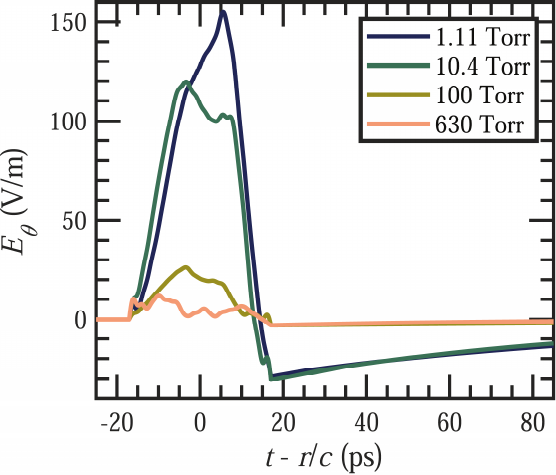}\end{figure}           

\textbf{Fig.~15} Far field $E_{\theta}$ for $\theta=15^{\circ}$ vs.\ $t$ for a
range of pressures $p$ with the constant current decay rates $\nu_{\max}$ for
each $p$ (Table~2) at spherical radius $r=1.2$ m.

\bigskip

To calculate the effect of the current decay rate on the far field, we express
measured $Q\left(  z\right)  $ as a series of sinusoidal oscillations with a
half-integer number of periods over the filament length, which we take to be
$L=30$~cm,
\begin{equation}%
\begin{tabular}
[c]{c}%
$Q\left(  z\right)  =%
{\displaystyle\sum\limits_{j=1}^{32}}
q_{j}\sin\left(  \frac{j\pi}{L}\left(  z-z_{0}\right)  +\frac{j\pi}{2}\right)
$\\
$z_{0}-\frac{L}{2}\leq z\leq z_{0}+\frac{L}{2}$ \ \ elsewhere \ $Q\left(
z\right)  =0$\\
$q_{j}=\frac{2}{L}\int_{z_{0}-\frac{L}{2}}^{z_{0}+\frac{L}{2}}\sin\left(
\frac{j\pi}{L}\left(  z-z_{0}\right)  +\frac{j\pi}{2}\right)  Q\left(
z\right)  dz$%
\end{tabular}
\ \label{X34}%
\end{equation}
$q_{j}$ is found by multiplying the sum by $\sin\left(  k\pi\left(
z-z_{0}\right)  /L+k\pi/2\right)  $ and integrating each term over the
filament domain (zero for $k\neq j$.) From Eq.~\ref{X33},%

\begin{equation}%
\begin{tabular}
[c]{c}%
$E_{0\theta}\left(  t\right)  =%
{\displaystyle\sum\limits_{j=1}^{32}}
\frac{q_{j}j\left(  \sin\theta-\sin2\theta\right)  \cos\left(  \frac
{j\pi\left(  ct-r\right)  }{L\eta}+\frac{j\pi}{2}\right)  }{4\epsilon_{0}%
L\eta^{3}r}$\\
$\frac{r}{c}-\frac{L\eta}{2c}\leq t\leq\frac{r}{c}+\frac{L\eta}{2c}$
\ \ \ $\eta=1-\cos\theta$%
\end{tabular}
\label{X35}%
\end{equation}

The effect of $\nu$ can be approximated by linear convolution over $t$ of
Eq.~\ref{X35} with normalized current waveform $\xi\left(  t\right)
=\Theta\left(  t\right)  \nu e^{-\nu t}$, where $\nu$ is constant. $\nu
=\nu_{1}$ does not vary strongly with $z$ (Fig.~14) and has several places
where the implied $Q$ is inconsistent with that directly measured (Fig.~13).
Two possible choices for $\nu$, then, are the values of $\nu_{1}$ and $\nu
_{0}$\ nearest to $z=z_{0}$~cm for which the two $Q$ estimates are equal
(Table~2). These are assumed to be where multiple filamentation is not so
severe as to invalidate the $I_{0}$, $n$, and $T$ results plotted in Fig.~11.
The contour plots of Fig.~6 provide a degree of corroboration.

The following convolution applied to Eqs.~\ref{X28} causes the (linearly
related) impulsive current assumed for $\mathbf{E}_{0}\left(  t\right)  $ to
be replaced by one with a rapid rise and exponential decay with the same time
integral $Q$,%

\begin{equation}%
\begin{tabular}
[c]{l}%
$E_{\theta}=E_{0\theta}\ast\xi\equiv\int_{-\infty}^{+\infty}E_{0\theta}\left(
t-\tau\right)  \xi\left(  \tau\right)  d\tau$\\
$\tau_{\pm}=t-\frac{r}{c}\pm\frac{L\eta}{2}$ \ \ \ \ \ \ \ \ $t\geq\frac{r}%
{c}-\frac{L\eta}{2c}$%
\end{tabular}
\ \ \ \ \label{X36}%
\end{equation}
We replace the upper and lower integration limits with $\tau_{\pm}$ since
$E_{0\theta}\left(  t-\tau\right)  =0$ outside of this domain (Eq.~\ref{X35}%
.1). To facilitate an analytic solution, $\Theta\left(  \tau\right)  $ is
removed from the the definition of $\xi\left(  \tau\right)  $ by replacing the
lower limit with $0$ for $t\leq\frac{r}{c}+\frac{L\eta}{2c}$. $\Theta\left(
\tau\right)  $ is already redundant for $t\geq\frac{r}{c}+\frac{L\eta}{2c}$.
This corresponds to the retarded time that the optical pulse has exited the
filament, but the current is still decaying. The solution is,%
\begin{equation}%
\begin{tabular}
[c]{l}%
for $\frac{r}{c}-\frac{L\eta}{2c}\leq t\leq\frac{r}{c}+\frac{L\eta}{2c}$,
$\ E_{\theta}\mathbf{=}%
{\displaystyle\sum\limits_{j=1}^{32}}
\frac{q_{j}j\nu\left(  \sin\theta-\sin2\theta\right)  }{4\epsilon_{0}\eta
^{2}r}$\\
$\times\frac{\nu L\eta\left(  \cos\left(  \frac{j\pi\left(  ct-r\right)
}{L\eta}+\frac{j\pi}{2}\right)  -\exp\left(  -\nu\frac{2\left(  ct-r\right)
+L\eta}{2c}\right)  \right)  +j\pi c\sin\left(  \frac{j\pi\left(  ct-r\right)
}{L\eta}+\frac{j\pi}{2}\right)  }{\left(  \nu L\eta\right)  ^{2}+\left(  j\pi
c\right)  ^{2}}\ $\\
for $t\geq\frac{r}{c}+\frac{L\eta}{2c}$, $E_{\theta}\mathbf{=}%
{\displaystyle\sum\limits_{j=1}^{32}}
\frac{q_{j}jL\nu^{2}\left(  \sin2\theta-\sin\theta\right)  }{4\epsilon_{0}\eta
r}$\\
$\times\frac{\exp\left(  -\frac{\nu L\eta}{2c}\right)  -\left(  -1\right)
^{j}\exp\left(  +\frac{\nu L\eta}{2c}\right)  }{\left(  \nu L\eta\right)
^{2}+\left(  j\pi c\right)  ^{2}}\exp\left(  -\nu\frac{ct-r}{c}\right)  $%
\end{tabular}
\ \ \ \ \label{X37}%
\end{equation}

Since $\nu_{1}$ (Eq.~\ref{X23}) and $\nu_{0}$ (Eq.~\ref{X27}) assume $U_{0}$
is only dissipated Ohmically, the radiated energy $K_{\text{R}}$ is not
accounted for. Upon integration of radiated power flux density\ $c\epsilon
_{0}E_{\theta}^{2}$\ of Eq.~\ref{X37} over $t$, and then over solid angles
from $\theta=\theta_{\min}$ to $60^{\circ}$ (beyond which it is negligible),
one finds that $K_{\text{R}}\rightarrow\infty$ as $\theta_{\min}\rightarrow0$.
This waveform, though, assumes an instantaneous current rise followed by the
Ohmic decay of a collisional plasma with real $\sigma$. This is unphysical for
time scales less than $\nu_{\text{c,eff}}^{-1}$. $\theta_{\min}$ is chosen,
then, to be that which, for a given $\nu$, leads to $f_{1}=\nu_{\text{c,eff}}$
at $\theta=\theta_{\min}$, where $f_{1}=c/\left(  2L(1-\cos\theta)\right)  $
(as already defined) is the principal frequency of the sinusoidal expansion of
Eqs.~\ref{X37}. \ The value of $\nu=\nu_{\max}$ that results in $K_{\text{R}}$
with this $\theta_{\min}$ equal to $K_{0}$, the latter defined as the integral
of $U_{0}=\lambda_{I}I_{\text{f}0}^{2}$ over $z$, corresponds to the full
conversion of laser-imparted center of mass kinetic energy to radiation with
$\theta\geq\theta_{\min}$. $\nu_{\max}$ is an upper bound on $\nu$ since less
than 100\% efficiency of conversion into such radiation implies an even
smaller $\nu$. To further assure this, we set $U_{0}=0$ for all $z$ locations
along the integral domain where it is undefined due to one of the parameters
it is a function of (Fig.~8 and Fig.~11) going off-scale or not being
measured. Figure~15 plots $E_{\theta}$ for $\nu=\nu_{\max}$. $r=1.2$~m is
assumed for comparison with published microwave antenna measurements taken
under similar conditions \cite{Englesbe18}, albeit with a laser pulse energy
of $40$~mJ vs.\ our $30$~mJ.%

\[%
\begin{tabular}
[c]{l|l|llll}%
$\nu$ (s$^{-1}$) & $\theta$ ($^{\circ}$) & $1.11$~Torr & $10.4$~Torr &
$100$~Torr & $630$~Torr\\\hline
$\nu_{1}$ & 15 & $142$ & $654$ & $736$ & $787$\\
$\nu_{0}$ & 15 & $201$ & $793$ & $762$ & $834$\\
$\nu_{\max}$ & 15 & $183$ & $150$ & $29.0$ & $14.8$\\
Eng. & 15 & \allowbreak130 & \allowbreak55 & 16 & \allowbreak4\\\hline
$\nu_{1}$ & 30 & $14.2$ & $42.8$ & $27.5$ & $27.7$\\
$\nu_{0}$ & 30 & $19.4$ & $47.0$ & $30.0$ & $28.5$\\
$\nu_{\max}$ & 30 & $17.9$ & $15.7$ & $3.04\ \ $ & $\ 1.49\ \ \ \ \ $\\
Eng. & 30 & \allowbreak26 & \allowbreak\allowbreak11 & \allowbreak
3.\thinspace\allowbreak2 & \allowbreak0.8\\\hline
$\nu_{1}$ & 45 & $2.14$ & $4.36$ & $2.95$ & $2.31$\\
$\nu_{0}$ & 45 & $2.77$ & $4.59$ & $3.10$ & $2.30$\\
$\nu_{\max}$ & 45 & $2.60$ & $2.33$ & $0.487$ & $0.237$\\
Eng. & 45 & \allowbreak10.\allowbreak4 & \allowbreak$\allowbreak$4.4 &
\allowbreak1.28 & \allowbreak0.32
\end{tabular}
\ \ \ \ \ \
\]

\textbf{Table~1} Columns 3-6 list far field $\mathbf{E}$ peak-peak amplitude
$E_{\theta\text{pp}}$ (V/m) at $r=1.3$~m for the decay rate $\nu$, emission
angle $\theta$, and pressure $p$ listed in Row 1. $\nu_{1}$ and $\nu_{0}$ are
solutions to Eq.~\ref{X23} and Eq.~\ref{X27}, respectively. $\nu_{\max}$ is
the $\nu$ value corresponding to 100\% conversion of the electron
center-of-mass kinetic energy imparted by the laser $K_{0}$ to radiated energy
$K_{\text{R}}$. Row 5 (labeled \textquotedblleft Eng.\textquotedblright) lists
$E_{\theta\text{pp}}$ reported by Englesbe \cite{Englesbe18}, based on voltage
peak-peak amplitudes of the upper trace of Fig.~E2a (referring to Fig.~2a of
Englesbe \cite{Englesbe18}) times the calibration constant plotted in Fig E1b
at our $\theta=15^{\circ}$ principal frequency of $f_{1}=14.7$~GHz
(Eqs.~\ref{X37}, $j=1$). Row 9 [13] lists Row 5 values multiplied by $0.2$
[$0.08$]. This is the ratio $0.8$ [$0.5$] of the polar plot amplitudes of
Fig.~E2b at $\theta=30^{\circ}$ [$45^{\circ}$] to those at\ $\theta=15^{\circ
}$ times the ratio $0.25$ [$0.17$] of the calibration factors plotted in
Fig.~E1b for principal frequency $f_{1}=3.7$~GHz [$1.72$~GHz] for
$\theta=30^{\circ}$ [$45^{\circ}$] to that at $f_{1}=14.7$~GHz for
$\theta=15^{\circ}$.

\bigskip%

\[%
\begin{tabular}
[c]{l|llll}%
Parameter & $1.11$~Torr & $10.4$~Torr & $100$~Torr & $630$~Torr\\\hline
$z_{1}$ (cm) & $323$ \  & $310$ & $313$ \ \  & $309$\\
$-Q$ (pC) & $0.494$ & $\allowbreak0.184$ & $0.607$ & $0.0554$\\
$\nu_{\text{c,eff}}$ (s$^{-1}$/10$^{9}$) & $8.69$ & $70.4$ & $778$ & $3367$\\
$\nu_{1}$ (s$^{-1}$/10$^{9}$) & $8.62$ & $54.8$ & $415$ & $2167$\\
$\nu_{0}$ (s$^{-1}$/10$^{9}$) & $12.1$ & $67.9$ & $500$ & $2537$\\
$\nu_{\max}$ (s$^{-1}$/10$^{9}$) & $11.05$ & $13.41$ & $6.63$ & $16.63$\\
$I_{\text{f}0}$ (mA) & $4.3$ & $144$ & $517$ & $217$\\
$I_{\text{f}}$ (mA) & $4.3$ & $101$ & $252$ & $120$\\
$K_{0}$ (nJ) & $0.0698\allowbreak$ & $1.11$ & $0.421$ & $0.360$\\
$\theta_{\min}$ ($^{\circ}$) & $19.5$ & $6.8$ & $2.05$ & $0.99$%
\end{tabular}
\ \ \
\]

\textbf{Table~2} Parameters are tabulated at axial locations $z=z_{1}$, where
our independent estimates of filament current time integral $Q$ from direct
measurement (Eq.~\ref{X16}) and resistive decay ($I_{\text{f}}$ from
Eq.~\ref{X26} divided by $\nu_{1}$ from Eq.~\ref{X23}) are equal (Fig.~13).
$I_{\text{f}0}$ is the post-optical filament current (Eq.~\ref{X17}) before
diversion of a portion of its center of mass kinetic energy per unit length
$U_{0}$ to a self-consistent internal EM field (Eqs.~\ref{X24}). $\nu_{0}$
(Eqs.~\ref{X27}) is a complementary decay rate based on the post-optical
current being driven to the filament periphery and diffusing back in.
$\theta_{\min}$ is the far field emission angle where $f_{1}=\nu
_{\text{c,eff}}$, below which our model is invalid. $\nu_{\max}$ (an upper
bound on $\nu$) is the decay rate implied by $Q$ assuming full conversion of
total center of mass electron kinetic energy $K_{0}$ to far field radiation
emitted at $\theta\geq\theta_{\min}$.

\section{Discussion and conclusions}

The primary purpose of this paper is to provide experimental results for use
as benchmarks for 3-D time domain simulations of the filamentation and
radiation process. The physics of interaction between the non-steady EM field
associated filament current in the optical pulse wake requiring $\nu$ values
well-below $\nu_{\text{c,eff}}$ to be consistent with our direct self-emission
based $Q\left(  z\right)  $ measurements is not addressed here. High values of
$E_{\theta\text{pp}}$ for $\nu_{0}$ and $\nu_{1}$ for all but the lowest
$p=1.1$ Torr (Table~1)\ imply $K_{\text{R}}\gg K_{0}$, so are deemed
unphysical. They are presented to emphasize the need for a non-steady state
mechanism lowering $\nu$ to account for the $Q$ measurements. Such long
current lifetimes have been directly measured \cite{Zhou11}. The $Q\left(
z\right)  $ measurement, which serves as the basis for Eq.~\ref{X37}, may be
considered accurate for values of $\nu$ as low as $\nu_{\max}$ since it is an
S-band diagnostic. $Q=I_{\text{f}}/\nu_{1}$ using Eq.~\ref{X23} and
Eq.~\ref{X26} being similar (within a factor of $\sim2$) to the self-emission
measurement (Fig.~13) suggests that the mechanism has little effect on $Q$
itself, such as a linear convolution like the one that results in $E_{\theta}$
(Eq.~\ref{X36}). This is, in part, why $z=z_{1}$ values (where the measured
$Q$ and $I_{\text{f}}/\nu_{1}$ agree) are retained as reference points for
properties with $\nu=\nu_{\max}$.

Table~1 shows that $\nu=\nu_{\max}$ results in a reasonably good
$E_{\theta\text{pp}}$ match with Englebe's data considering that $\nu_{\max}$
is only an upper bound. The following issues may account for the generally
lower experimental $E_{\theta\text{pp}}$: $K_{0}$ may be diverted to other
processes, such as Ohmic heating and $\theta<\theta_{\min}$ radiation. Re the
latter, the simple cutoff of $E_{\theta\text{pp}}$ at $\theta=\theta_{\min}$
for our $K_{\text{R}}$ integration is a crude way to enforce our collisional
plasma assumption. While our model assumes the $U_{0}$ source for
$I_{\text{f}t}$ is due exclusively to parent ion electron restrike-enhanced
$\mathbf{J}\times\mathbf{B}$ force during optical pulse passage
\cite{Ruden25QmodT}, Zhou, et al.~\cite{Zhou11} proposes an opposing
contribution to $I_{\text{f}t}$ resulting from space charge separation caused
by transverse pondermotive electron drift resulting from the optical
$\nabla\left\vert \mathbf{E}\right\vert ^{2}$. The models for $W_{i}\left(
I\right)  $, $T\left(  I_{0}\right)  $ and $\left\langle p_{{z}}\right\rangle
$ vs.\ $I_{0}$ needed to calculate $K_{0}$ have their own degrees of
uncertainty. The isolated primary filament assumption is not valid in many $z$
locations, while $K_{0}$ (as a collective property) assumes it is. The uniform
cylinder approximation used to represent a Gaussian distribution to calibrate
the S-band instrument is imprecise. The horns used to measure $E_{\theta}$ are
dispersive with a frequency-dependent sensitivity, which alters the signal
waveform, potentially lowering $E_{\theta\text{pp}}$. Reflection and
refraction through the quartz tube has a $\sim10\%$ effect in terms of
reducing the experimental $E_{\theta\text{pp}}$ at $\theta=15^{\circ}$.

One additional issue that may reduce $I_{\text{f}t}$ (and therefore
$E_{\theta\text{pp}}$) is Cherenkov radiation within the quartz tube wall due
to the speed of light within it being less than $c$. $E_{z}=-Q/\left(
2\pi\epsilon_{0}R_{1}^{2}\right)  $ for a point particle traveling at speed
near $c$ on-axis \cite{Baturin14}$^{:8}$. This is applicable to our impulsive
model since Eqs.~\ref{X29} refers to the spatiotemporal dependence of
point-like charge distribution, regardless of its nature; it need not be an
actual particle. It is more likely, though, to significantly reduce
$I_{\text{f}t}$ for our $R_{1}=3.92$~mm tube than\ the larger tube used by
Englesbe ($R_{1}\approx17$ mm, $R_{2}=19$~mm, without the small tube
transition seen in Fig.~1).

One noteworthy result is that, despite $K_{0}$ being $5$ times higher for
$p=630$ Torr than for $p=1.11$~Torr (Table~2), $E_{\theta\text{pp}}$ is at
least an order of magnitude lower than measured and for $\nu=\nu_{\max}$
(Table~1). This is due the rapid increase in $E_{\theta}$ with decreasing
$\theta$ (Eqs.~\ref{X37}) implying that radiated energy is highest near
$\theta_{\min}$. $\theta=15^{\circ}$, for example, is near $\theta_{\min
}=19.5^{\circ}$ for $p=1.11$~Torr, so radiation is strong at this angle.
However, $f_{1}=3.3$ THz at $\theta_{\min}=0.99^{\circ}$ for $p=630$~Torr, so
$f_{1}=14.7$~GHz at $\theta=15^{\circ}$ is at the tail end of the power spectrum.

Comparison of Fig.~15 to Fig.~E3a suggests the former represents a baseline
response of the far field to the wake field current that explicitly neglects
the oscillatory modes recorded by the latter. Englesbe's far field data
\cite{Englesbe18} is broadly consistent with Eqs.~\ref{X37} with $\nu
=\nu_{\max}$ by several metrics beyond those of Table~1. The initial fast rise
and fall to zero at $\theta=15^{\circ}$ for all $p$ seen in Fig.~E3a takes
$33$~ps, very similar to that seen in our Fig.~15. The greater undershoot (aka
\textquotedblleft ring\textquotedblright) of the former and slow oscillation
during subsequent decay not being present in our model may be attributed to
our neglect of oscillatory current modes (such as those recorded by Zhou
\cite{Zhou11}). Horn dispersion may also contribute to the oscillation. Also,
like Eq.~\ref{X33}, Fig.~E2c, shows a drop to very low $E_{\theta}$ for
$\theta>60^{\circ}$. Regarding the FFT-based results presented, the wide band
response of Fig.~E3C ($\theta=15^{\circ}$) is consistent with Fig.~15 ($f_{1}$
is near the middle in the FFT spectrum). A similar correlation is apparent for
other angles too. $f_{1}=3$, $6$ and $9$~GHz occur at $\theta=33.6^{\circ}$,
$23.6^{\circ}$, and $19.2^{\circ}$, respectively, in our model. These are
comparable to the mean $\theta$ of the $E_{\theta}$ FFT distribution
$\mathcal{F}_{E}\left(  \theta\right)  $ plotted in Fig.~E5 for these
frequencies: $\left\langle \theta\right\rangle =31.1\pm1.5^{\circ}$,
$22.3\pm0.6^{\circ}$, and $20.6\pm2.6^{\circ}$, respectively, where $\pm$
refers to the standard deviations of the spread in $\left\langle
\theta\right\rangle $ for the $3$ $p$'s of the plots.

The degree to which the laser repetition rate of $f_{\text{r}}=10$~s$^{-1}$
affects our ability to treat the filament as an isolated event is a worthwhile
consideration. The electron recombination time \cite{Papeer13} is $\ll100$~ns
(consistent with our Fig.~2) and the gas density profile due to the resulting
pressure wave returns to ambient \cite{Cheng13} in $\sim1$~ms, so these are
not significant issues. Empirically, energy deposition in a similar experiment
\cite{Pena24} exhibits single-shot behavior below $f_{\text{r}}=100$~s$^{-1}$.
However, this experiment does not use a small central quartz tube. Electron
recombination with O$_{2}^{+}$ often leads to molecular dissociation
\cite{Talebpour96}, potentially resulting in two instances of O $+$
O$_{2}\rightarrow$ O$_{3}$ per O$_{2}$ ionization \cite{Chapman30}. The time
required for the fraction of O$_{2}$ molecules to experience an ionization
event within the tube to reach $\eta_{\text{f}}$ $=0.1$ is $t_{\text{O}}%
=\eta_{\text{f}}R_{1}^{2}/\left(  f_{\text{r}}\eta_{R1}R^{2}\right)  $, where
$\eta_{R1}$ is the ionization fraction of O$_{2}$ for one shot in the uniform
cylinder representation. From Fig.~11a, Fig.~9, and Eq.~\ref{X17}.2, there is
a peak of $\eta_{R1}=3.2\times10^{-3}$ for $p=630$~Torr at $z=299$~cm where,
from Fig.~7 and Eq.~\ref{X9}, $R=56$~$\mu$m. So, $t_{\text{O}}=15,227$~s
($254$~min) there. The 1-D linear diffusion time \cite{Crank75}$^{:2.7}$ of
O$_{3}$ from the center of the tube of half-length $s_{\text{T}}=20$ cm to
either end, though, is only $t_{D}=s_{\text{T}}^{2}/\left(  2D\right)
=\allowbreak1333$ s ($\allowbreak22$ min), where $D=0.15\times10^{-4}$~m$^{2}%
$s$^{-1}$ is the O$_{3}$ diffusion coefficient in $1$~atm air
\cite{Langenberg20}. O$_{3}$ accumulation is slightly more significant for
$p=100$ Torr at its $\eta_{R1}$ peak ($t_{\text{O}}=26$~min vs.\ $t_{D}%
=3$~min), but with $D$ being inversely proportional to $p$, O$_{3}$ is not an
issue for the lower $p$ values either. The above neglects convection and the
finite lifetime of O$_{3}$. O$_{3}$ is not expected to be a significant effect
for our experiment.

\appendix

\section{Diagnostic calibration simulations}

\subsection{Geometry}%

\begin{figure}[H]\includegraphics[width=3.375in]{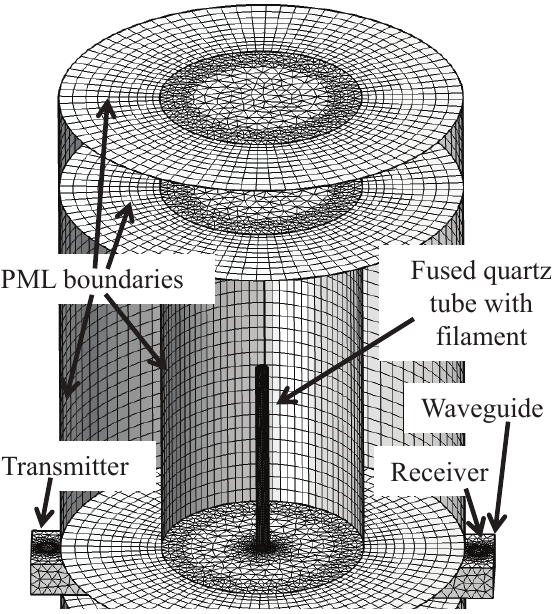}\end{figure}           

\textbf{Fig.~16} Upper portion of simulation geometry and mesh for 3-D
continuous guided wave attenuation by a conducting filament. This view
emphasizes the upper Perfectly Matched Layer needed to reduce feedback
affecting attenuation.

\bigskip

The combined effect of\ a wide range of $\sigma$ and $R$ values of a uniform
cylindrical filament within a fused quartz tube traversing an S-band waveguide
in the direction of the TE$_{10}$ mode's electric field on the attenuation of
a transmitter-driven $3.2$~GHz wave with that mode is simulated by 3-D
continuous wave COMSOL Multiphysics\texttrademark\ electromagnetic (EM)
simulations. The simulated power attenuation factor\ is,%
\begin{equation}
A\equiv\left(  \frac{V_{\text{R}}}{V_{\text{R0}}}\right)  ^{2}\text{
\ \ (theoretical)} \label{A1}%
\end{equation}
where $V_{\text{R}}$ and $V_{\text{R0}}$ are the simulated output adapter RMS
voltage of Eq.~\ref{X1} with and without the filament present, respectively.

\begin{figure}[H]\includegraphics[width=3.375in]{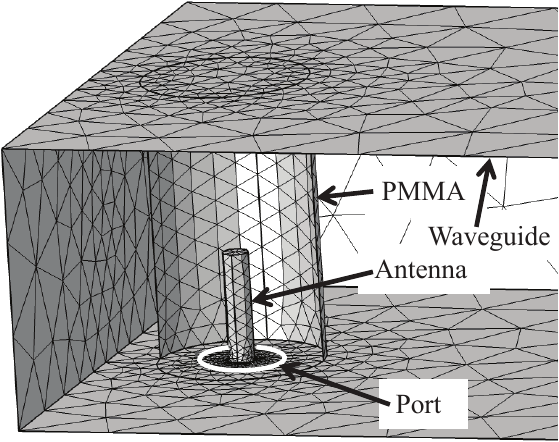}\end{figure}           

\textbf{Fig.~17 }Transmitter delivering $3.2$~GHz signal to waveguide via $50$
$\Omega$ port (return conductor outlined in white). The receiver at the other
end the waveguide is its mirror image.%

\begin{figure}[H]\includegraphics[width=3.375in]{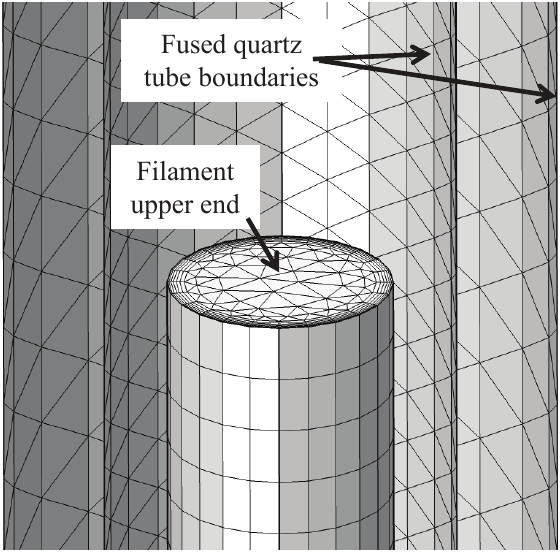}\end{figure}           

\textbf{Fig.~18} View of the top of $40$~cm long model filament (shown here
within the quartz tube) illustrating the axially swept mesh used, and its
concentration across the current skin depth. Filament radius is $R=2.5$~mm and
electrical conductivity is $\sigma=5045$ S/m in this example.

\bigskip

Figure~16 shows the upper portion of simulation geometry showing the portion
of the quartz tube containing the filament external to the waveguide. It is
surrounded by a Perfectly Matched Layer (PML) \cite{Jin10} needed to eliminate
radiation feedback affecting attenuation parameter $A$. The lower portion of
the PML enclosed external filament region is its mirror image. To better
represent the experimental filament's conductivity gradually tapering off
beyond at both ends, and thereby minimize unphysical reflected current from
the filament ends back into the waveguide for high conductivities
(irrespective of the PML), $G=\pi R^{2}\sigma$ is varied cosinusoidally from
the center in the simulation,%
\begin{equation}
G_{\text{actual}}\left(  z\right)  =G\cos\left(  \frac{\pi z}{H}\right)
\label{A2}%
\end{equation}
where $H=40$~cm is filament length.

The bottom annular boundary between the tight fitting hole that passes the
quartz tube through the waveguide and the outer PML boundary is perfectly
conducting and coplanar with the outer waveguide wall. The gap between the
waveguide inner boundary and the PML (ie., the depth of the hole) is equated
to the wall thickness of the experimental waveguide ($2$~mm). The lower
external filament and PML is the mirror image. The maximum mesh spacing is set
to be $1/10$'th of the $3.2$~GHz drive wavelength. It is much finer for small
components and resistive skin depths to accommodate the small scale-length of
field variations.%

Figure~17 is an expanded view of the simulated transmitter, which matches that
of the experiment, delivering $3.2$~GHz signal to the waveguide via $50$
$\Omega$ port (return conductor outlined in white). The receiver at the other
end is its mirror image. The simulation mesh is tetrahedral, except for the
PML and filament, as described.

Figure~18 shows the top of $40$~cm long model filament (shown here within the
quartz tube) illustrating the axially swept mesh used for a filament with
radius $R=10^{3.4}$~$\mu$m = $2.5$~mm and electrical conductivity
$\sigma=5045$~S/m. The outer layer has $6$ layers of swept quadrilaterals
decreasing in radial thickness with radius, and with a total thickness of the
lesser of three resistive skin depths $\delta_{0}$ or $R/2$. $\delta_{0}=R/20$
in this example. Given such a current confined to near the surface, the effect
on the $3.2$~GHz traveling wave is largely inductive. The inner filament
region transitions to a mesh of axially swept triangles.

\subsection{Filament unit length conductance $G$ vs.\ attenuation $A$ and
radius $R$}

Figure~19 plots $B_{x}$ for the case of (referring to Fig.~20 legend)
$R=10^{3.4}$~$\mu$m and $G=10^{-3}$~Sm. The EM waves beyond the filament are
strongly attenuated in the main simulations for which the calibration is based
due to the cosine tapering of $G$, by design. So, an axially uniform $G$ is
used for this plot to better show the efficacy of the PML in reducing reflections.

Figure~20 plots simulated $A$ vs.\ $G$ for a range of $R$ values out to the
tube inner radius. The legend identifies $\log R\left(  \mu\text{m}\right)  $
assumed for each curve. The insensitivity to $R$ for small $G$ shown in the
plots implies $A$ is dominated by resistive impedance. Conversely,\ the
saturation of $A$ for large $G$ plotted shows that inductive impedance
dominates as the current skin depth $\delta_{0}=\left(  \pi f_{0}\mu_{0}%
\sigma\right)  ^{-1/2}$ becomes small relative to $R$.

Figure~21 displays the calibration table for cosine tapered $G$ vs.\ $A$ and
$R$ as a contour plot. It is generated by linear interpolation in parameter
$\log\left(  R\text{ }\left(  \mu\text{m}\right)  \right)  $ between the
continuous curves plotted in Fig.~20 (which are already interpolated between
simulated values of $G$ for each $R$).%

\begin{figure}[H]\includegraphics[width=3.375in]{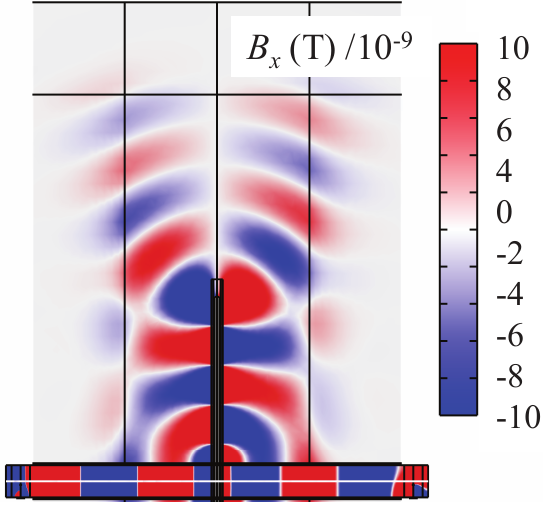}\end{figure}           

\textbf{Fig.~19} Saturated color map (upper portion) of the $x$ component of
the magnetic field $B_{x}$ (into the page) in the $y$-$z$ plane resulting from
the $3.2$~GHz drive for the high conductivity filament illustrated in Fig.~18
with an axially uniform $G=\pi R^{2}\sigma=10^{-3}$~Sm. It best shows the
effect of the Perfectly Matched Layer (PML) in preventing reflections that
would otherwise feed back to the filament current, perturbing the value of
attenuation parameter $A$ relative to a filament extending into free space.%

\begin{figure}[H]\includegraphics[width=3.375in]{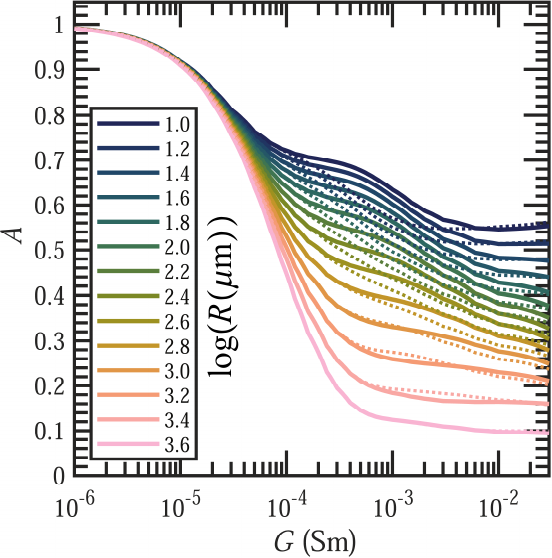}\end{figure}           

\textbf{Fig.~20} Receiver power attenuation factor $A$ vs.\ $G$ for a range of
$R$ specified by its base 10 log in the legend. The continuous curves are a
piecewise linear fit interpolated between $A$ vs.\ $G$ values calculated from
numerous simulations over a range of cosine tapered $G$. The dotted lines plot
the results for simulations with axially uniform $G$ for comparison.%

\begin{figure}[H]\includegraphics[width=3.375in]{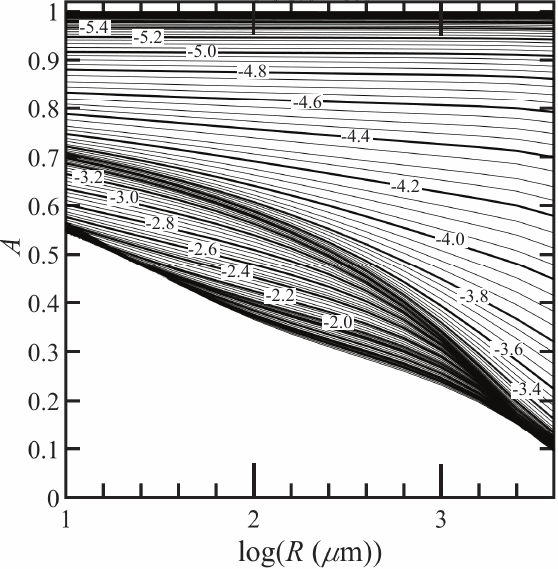}\end{figure}           

\textbf{Fig.~21 }Contour plot of cosine tapered $G\left(  A,R\right)  $
calibration table showing the central $\log\left(  G\text{ }\left(
\text{Sm}\right)  \right)  $ (major line labels) vs.\ $A$ and $\log\left(
R\text{ }\left(  \mu\text{m}\right)  \right)  $.

\subsection{Validation}%

\begin{figure}[H]\includegraphics[width=3.375in]{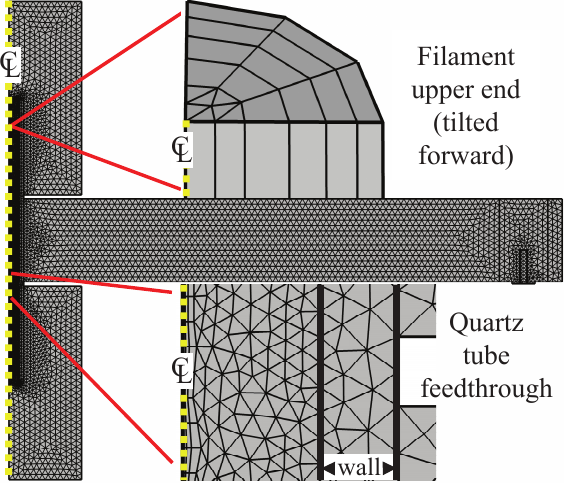}\end{figure}           

\textbf{Fig.~22 }Geometry and mesh used for 3-D time-domain EM simulations of
filament self-emission in the $y$-$z$ plane. The details shown are of the end
of the filament itself (with reflecting $y$-$z$ and $x$-$z$ planes and
surrounding mesh omitted) and the lower feed through which the quartz tube
containing the filament passes. For scale, the filament radius is $25.1$~$\mu
$m, the tube outer radius is $6.15$~mm, and the waveguide half-length is
$20.65$~cm.

\bigskip

$A$ is measured as the ratio of the shot-averaged HP detector voltage
(Eq.~\ref{X1}) corrected for self-emission with ($V_{\text{S}}$) and without
($V_{\text{S0}}$) a filament present,%
\begin{equation}
A=\frac{V_{\text{S}}}{V_{\text{S0}}}\text{ \ \ (experimental)}\label{A3}%
\end{equation}
The $G\left(  A,R\right)  $\ calibration table plotted in Fig.~21 is validated
by comparing the theoretical\ values of $A$ for several cases of actual
uniform solid conducting filaments with measured values of $R$ and $\sigma$ of
the same length as modeled ($40$~cm) with experimental measurements of $A$
with these filaments inserted into the waveguide in lieu of the plasma
filament. These simulations, therefore, are performed with an axially uniform
$G$, and a corresponding calibration table is created for it.
Proto-pasta\texttrademark\ and BlackMagic3D\texttrademark\ electrically
conducting 3-D printer filament stock are either used directly ($R\approx
0.875$~mm) or extruded via a 3-D printer's automatic feed initialization mode,
and annealed straight in boiled water. These materials are rated to have
$\sigma$ values of roughly $100$~S/m and $6.67$~S/m, respectively. After
measuring $R$ with calipers, actual $\sigma$ is determined by based on the
rate of change of resistance with distance between contacts with an Ohm meter.
To verify $A$ saturation values in the high $G$ limit (where attenuation is
dominated by inductance), Cu, SS316L stainless steel, and W wires with various
$R$ are also tested. The results are presented in Table~3. The tabulated
values of $\sigma$ for Cu and W are standard cited values. However, the skin
layer for them proved too thin to simulate (a mesh could not be made), so
perfectly conducting boundary conditions for those wires were used to
determine the tabulated theoretical values.%

\[%
\begin{tabular}
[c]{cccccc}%
Material & $R$ (mm) & $\sigma$ (S/m) & $G$ (Sm) & $A$ (exp.) & $A$
(theor.)\\\hline
PP & $0.181\,$ & $24.4$ & $2.\,\allowbreak52$E$-6$ & $0.969$ & $0.979$\\
BM & $\,0.193$ & $197.$ & $2.\,\allowbreak29$E$-5$ & $\,0.847\,$ & $0.828$\\
PP & $0.875$ & $19.0$ & $4.\,\allowbreak57$E$-5$ & $0.603\,$ & $0.698$\\
BM & $0.865$ & $174.$ & $4.\,\allowbreak09$E$-4$ & $0.340\,$ & $0.377$\\
SS316L & $0.040$ & $1.35$E$6$ & $6.\,\allowbreak75$E$-3$ & $0.421$ & $0.449$\\
Cu & $0.10$ & $5.96$E$7$ & $1.87$ & $0.343$ & $0.360$\\
W & $2.38$ & $1.79$E$7$ & $317.\,$ & $0.137$ & $0.141$%
\end{tabular}
\]

\textbf{Table~3} Proto-pasta\texttrademark\ (PP in Row 1),
BlackMagic3D\texttrademark\ (BM), SS316L, copper, and tungsten wire validation
test results. Experimentally measured attenuation parameter $A$ results are
compared to interpolated COMSOL continuous wave simulation results (Fig.~20
dotted lines) for a range of wire radii $R$ and uniform conductivity $\sigma$.

\subsection{Self-emission simulation}%

\begin{figure}[H]\includegraphics[width=3.375in]{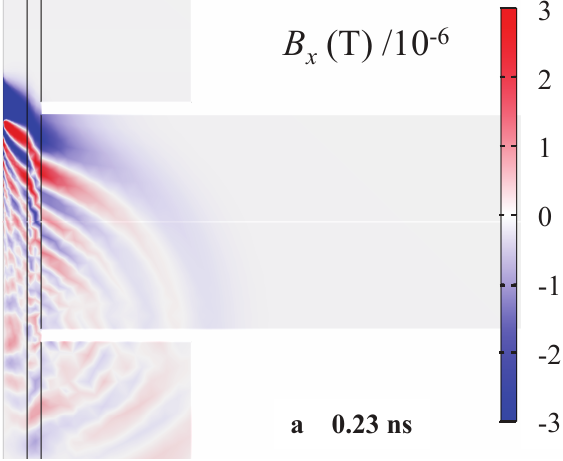}\end{figure}           
\begin{figure}[H]\includegraphics[width=3.375in]{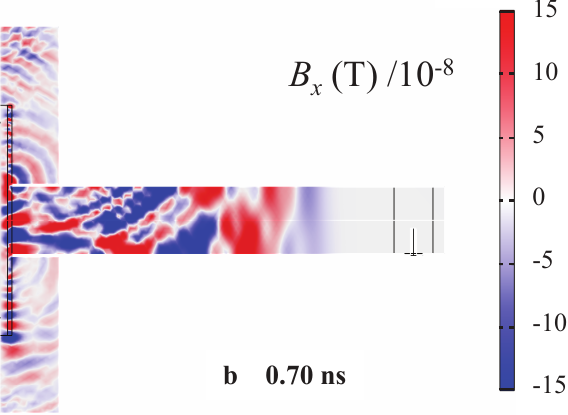}\end{figure}           

\textbf{Fig.~23 }$B_{x}$ in the $y$-$z$ plane at (a) $t=0.23$~ns and (b)
$t=0.70$~ns using a Gaussian current waveform with a standard deviation of $5$~ps
and a time integral of $-1$~pC for $I_{0}\left(  t\right)  $.

\bigskip

3-D full time-domain EM simulations of the effect of an axial current pulse
following the USPL laser pulse on the $50$~$\Omega$ receiver's voltage output
were performed to better understand the nature of self-emission. Various
filament radii, trailing conductivities, and current waveforms $I_{0}\left(
t\right)  $ were modeled which follows laser pulse passage. The filament
current at axial distance $z$ from the waveguide center is then,%
\begin{equation}%
\begin{tabular}
[c]{l}%
$I\left(  z\text{,}t\right)  =I_{0}\left(  t-\frac{z+0.5H}{c}\right)
\cos\left(  \frac{\pi z}{H}\right)  $\\
$\text{ for }-\frac{H}{2}\leq z\leq+\frac{H}{2}\text{ and }t\geq0$%
\end{tabular}
\ \ \ \ \ \ \ \ \ \ \ \label{A4}%
\end{equation}
where $H=10$~cm is the filament length now. The cosine term is included so
that the current pulse amplitude gradually rises from and back to zero,
peaking within the waveguide to avoid an intense unphysical EM pulse due to a
sudden current change. Current is applied only to the filament \emph{surface}
due to code limitations; distributed beams are not supported. However, the
electrical conductivity is distributed uniformly throughout the filament
volume. This results in unphysical high-frequency artifacts if both are
present at the same place and time, so the conductivity is only turned on to
$G=10^{-4}$~Sm at $t=1$ ns, when the EM disturbance first reaches the receiver.

The mesh, as illustrated in Fig.~22, has a maximum size of 2.5~mm in these
runs ($1/6$ times the vacuum wavelength of 20~GHz microwaves) to represent
high order modes. It is finer near the filament to resolve the current. A
swept quadrilateral boundary layer exterior to the filament is added to
mitigate this. The filament is shortened to $1/4$ the length of the
calibration simulation filaments, and the extensive PML is replaced with a
much smaller Absorbing Boundary Condition (ABC) \cite{Jin10}. There is no
longer a $3.2$~GHz drive, so symmetry is exploited by imposing perfect
magnetic conducting boundary conditions at the $x-y$ and $x-z$ planes.
Figure~23 displays the results of the simulation just before the current pulse
exits the waveguide, and at a later time when the EM disturbance has almost
reached the receiver.

\smallskip

\textbf{Funding. \ }This material is based on work supported by Air Force
Office of Scientific Research award FA9550-19RDCOR027.

\textbf{Acknowledgment. \ }The authors thank Travis Garrett for useful conversations.

\textbf{Disclosures. \ }The authors declare no conflicts of interest.

\textbf{Disclaimer. \ }The views expressed are those of the author and do not
necessarily reflect the official policy or position of the Department of the
Air Force, the Department of Defense, or the U.S. government.

\textbf{Data availability. \ }Data underlying the results presented in this
paper are available upon request of the primary author.

\textbf{Release approval. \ }Approved for public release; distribution is
unlimited. Public Affairs release approval \#AFRL-2026-3155

\newif\ifabfull\abfulltrue

\end{document}